\documentclass[preprint,aps,pra,showpacs,floatfix]{revtex4}
\usepackage{graphicx}
\usepackage{times}
\usepackage {euscript}
\usepackage{amsthm}
\usepackage{epsf}
\usepackage{epsfig}
\usepackage{bm}
\usepackage{bbm}
\renewcommand{\vec}[1]{{\mbox{\boldmath$#1$}}}
\newcommand{\vecsm}[1]{{\mbox{\scriptsize \boldmath$#1$}}}
\newcommand{\sumpr}{\mathop{{\sum}'}}
\begin{document}
%
\title{Relativistic calculations of the
charge-transfer probabilities and cross sections for low-energy
collisions of H-like ions with bare nuclei}
\author{I.~I.~Tupitsyn,$^{1}$ Y.~S.~Kozhedub,$^{1}$ V.~M.~Shabaev,$^{1}$
G.~B.~Deyneka,$^{2}$ S.~Hagmann,$^{3}$ C.~Kozhuharov,$^{3}$ G.~Plunien,$^{4}$
and
Th.~St\"ohlker$^{3,5,6}$}
\affiliation{
$^1$ Department of Physics, St. Petersburg State University,
Ulianovskaya 1, Petrodvorets, 198504 St. Petersburg, Russia \\
$^2$ St. Petersburg State University of Information Technologies,
Mechanics and Optics, Kronverk av. 49, 197101 St. Petersburg,  
Russia \\
$^3$
GSI Helmholtzzentrum f\"ur Schwerionenforschung GmbH,
Planckstrasse 1, D-64291 Darmstadt, Germany \\
$^4$ Institut f\"ur Theoretische Physik, Technische Universit\"at Dresden,
Mommsenstra{\ss}e 13, D-01062 Dresden, Germany\\
$^5$ Physikalisches Institut, Universit\"at Heidelberg,
Philosophenweg 12, D-69120 Heidelberg, Germany \\
$^6$Helmholtz-Institut Jena, D-07743 Jena, Germany
}
%
\begin{abstract}

A new method for solving the time-dependent two-center
Dirac equation is developed. The time-dependent Dirac wave function
is represented as a sum of atomic-like Dirac-Sturm orbitals,
localized at the ions. 
The atomic orbitals are obtained by solving
numerically the finite-difference one-center Dirac and
Dirac-Sturm equations with the potential which
is the sum of the exact reference-nucleus potential and
a monopole-approximation potential from the other nucleus.
 An original procedure to calculate
the two-center integrals with these orbitals is proposed.
The approach is tested by calculations of the charge transfer and
ionization cross sections for the H(1s)--proton collisions at proton energies
from 1 keV to 100 keV.
The obtained results are compared
with related experimental and other theoretical data.
To investigate the role of the relativistic effects, the charge
transfer cross sections for the Ne$^{9+}$($1s$)--Ne$^{10+}$ (at energies from
$0.1$ to $10$ MeV/u)
and U$^{91+}$(1s)--U$^{92+}$ (at energies from 6 to 10 MeV/u) collisions
are calculated in both relativistic and nonrelativistic cases.
\end{abstract}

\maketitle
\section{Introduction}
Since the pioneering works \cite{Firsov_51, Demkov_52, Bates_53},
where the oscillatory behavior of the resonance charge-transfer probability
for low-energy collisions was predicted, numerous publications have been devoted
to the theoretical investigations of the charge-transfer, excitations
and ionization in the H(1s)-H$^{+}$ collisions (see, e.g., reviews
\cite{Bransden_92, Fritsch_91, Winter_05}).
Nonrelativistic two-center finite basis set calculations
 have been carried out in Refs.
\cite{Gallaher_68, Shakeshaft_76, Reading_81, Fritsch_83, Ermolaev_90,
Toshima_99, Winter_09}.
Nonrelativistic three-dimensional lattice methods in the
position and momentum
spaces have been applied for the time-dependent Schr\"odinger equation in
Refs. \cite{Grun_82, Kolakowska_98, Kolakowska_99, Schultz_99, Tong_00}.
Within the nonrelativistic approach, 
 the probabilities and cross-sections for
a homonuclear collision
A$^{(Z-1)+}(1s)-$A$^{Z+}$ for the nuclear charge $Z>1$
can be easily obtained by scaling to 
the H($1s$)$-$H$^{+}$ collision. In the straight-line
trajectory approximation, the cross section $\sigma(Z,v)$ scales exactly as
$\sigma(Z,v)=\sigma(1,v/Z)/Z^2$ \cite{Briggs_73, Bransden_92},
where $v$ is the projectile velocity. This scaling law is not valid,
however, in the relativistic theory.

Collisions involving highly charged ions provide
tests of relativistic and quantum electrodynamics
 effects in the scattering theory  
\cite{Eichler_95,Shabaev_02,Eichler_07}. The study of such
processes can provide also a unique tool to probe
the quantum electrodynamics (QED) in the supercritical
Coulomb field, if the total charge of the colliding ions $Z=Z_A+Z_B$ is
larger than the critical one $Z_c=173$ (see, e.g., Refs. \cite{Zeldovich_71,
Rafelski_71, Greiner_85, Muller_94} and references therein). In the
presence of such a field the energy of the one-electron $1\sigma_{+}$ state
 of the quasi-molecule can reach the negative-energy Dirac
 continuum, when the distance $R$ between target ion $A$ and
 projectile ion $B$ becomes equal to the critical value $R_{\rm c}$. For the
 distance $R$ less than $R_{\rm c}$ the ground state level dives
into the negative-continuum spectrum. In the U$^{91+}$(1s)-U$^{92+}$
collision the critical radius for the point nucleus case was found to be 
$R_{\rm c}=36.8$ fm \cite{Rafelski_76}.

To date various approaches were developed to treat 
the heavy-ion collisions \cite {Eichler_05}.
In Refs. \cite{Becker_86, Strayer_90, Thiel_92,Momberger_96,
Wells_96, Ionescu_99, Pindzola_00, Busic_04},
the two- and three-dimensional numerical lattice methods were
employed to solve the time-dependent Dirac equation
at high energies. In Refs.  \cite{Eichler_90,Rumrich_93, Momberger_93,Gail_03},
high energy relativistic collisions of heavy ions were
considered using  the basis set approach, in which
 the time-dependent wave
function was expanded in terms of the atomic eigenstates of the
projectile and the target. For internuclear distances 
smaller than about  1000 fm some effects can also be evaluated  
within  so-called monopole approximation, which accounts only
for the spherically-symmetric part of the two-center
potential \cite{Soff_78, Reus_84, Ackad_08}.
The atomic processes such as excitation, ionization and charge transfer
in relativistic atomic collisions involving heavy and highly-charged
projectile ions with energies ranging from 100 MeV/u upward were studied
in Refs. \cite{Eichler_95,Stohlker_98a, Stohlker_98b, Ionescu_99,Ionescu_03,
Busic_04} and references therein.

In the present work, we develop a new method for solving
the two-center stationary and  time-dependent  
Dirac equations. 
The wave functions are expanded
in terms of the  Dirac and Dirac-Sturm basis functions,
which are central-field 4-component Dirac bispinors centered at the ions.
The radial parts of these orbitals are obtained by solving numerically
the finite-difference radial one-center Dirac and Dirac-Sturm equations.
In the nonrelativistic calculations of atoms and molecules,
so-called Coulomb-Sturmian basis set  was
introduced in Ref. \cite{Rotenberg_70}.
The Hartree-Fock calculations of atoms with this basis
were considered by many authors
(see, e.g., Ref. \cite{Gruzdev_87}).
The relativistic Coulomb-Sturmian basis was employed
in the papers \cite{Manakov_73, Drake_88, Avery_98, Grant_00}.
In the present paper we use a non-Coulomb relativistic Sturm basis set,
which is obtained by solving numerically the Dirac-Sturm equations with
a special choice of the weight function, that was proposed in Refs.
\cite{Tupitsyn_03, Tupitsyn_05}.
This allows us  to include any central-field potential in the
radial equations for the large and small components of the basis
functions. In particular, the Coulomb potential of the other ion can be included
in the radial equations within the monopole approximation.
The basis set constructed in this way
is described in detail in section \ref{subsec:basis}.

Calculations of two-center integrals
with the basis functions obtained require using special tools.
In the nonrelativistic case, a special symmetrical procedure for
such calculations, based on the
L\"owdin reexpansion \cite{Lowdin_56}, was developed in Refs.
\cite{Kotochigova_95, Tupitsyn_98}. 
In section \ref{subsec:integrals},
we generalize this procedure to the relativistic case.

To test the quality of the two-center expansion described above
 we perform relativistic calculations of the 
ground-state energy of molecular ion H$_2^{+}$ and one-electron
quasi-molecule Th$_2^{179+}$ at the ``chemical'' distance $R=2/Z$ a.u.
and compare the results with high-precision calculations
of Refs. \cite{Yang_91, Kullie_01}. We also calculate the
ground-state energy  as a function
of the internuclear distance $R$ and the critical radii $R_{\rm c}$ for a number
of
one-electron quasi-molecules, including U$_2^{183+}$. Most calculations
of the critical distances $R_c$ presented in the literature were performed
either for  the point-nucleus model \cite{Rafelski_76, Lisin_77, Matveev_00} or
with a crude estimate of the nuclear-size effect
\cite{Muller_76, Lisin_80, Popov_01}.
We calculate the critical distances
 for both point and extended nucleus models
 using the same basis set expansion.
The obtained results and comparison with 
the calculations by other authors are presented in section
\ref{subsec:stationary}.

The classical Rutherford trajectories \cite{Greiner_85}
of the projectile and target ions are obtained by numerical
solution of the Newton's equations. The
Born-Oppenheimer approximation is used to separate the motion of
the electron and the nuclei.
The magnetic interaction 
between the electron and the moving ions is neglected, because of low velocity
of the projectile with respect to the target.
The time-dependent Dirac equation for the electron is
solved using the two-center basis set expansion. 
The expansion coefficients can be defined employing, 
e.g., the Crank-Nicholsen propagation scheme \cite{Crank_47}
or the split-operator method \cite{Fett_82}. These methods
conserve the norm of the time-dependent wave function at each time step,
since the Crank-Nicholsen operator and the split-operator are unitary.
However, in this work we use the direct evolution exponential operator
method, which is more stable compared to the others. 
To obtain the matrix of the exponential operator in the finite
basis set one has to diagonalize
the generalized Hamiltonian matrix at each step of time.
Since our basis set is not too large, the diagonalization procedure
is not too time consuming.
The amplitudes of the charge transfer to different bound states
of the projectile ion are calculated nonperturbatively by projecting
the time-dependent wave function onto the moving Dirac orbitals
of the projectile. 

In section \ref{subsec:Charge-trnasfer} we present the results of the
relativistic calculations of the charge-transfer probabilities
and cross sections for the H(1s)--H$^{+}$, Ne$^{9+}$($1s$)--Ne$^{10+}$,
Xe$^{53+}$($1s$)--Xe$^{54+}$, and U$^{91+}$($1s$)--U$^{92+}$
low-energy collisions.
All the calculations are performed in the laboratory frame $S$,
that is defined to be at rest with respect to
the initial target position.
 The H($1s$)--H$^{+}$ collision is considered
in  section \ref{subsec:H2}. Since the relativistic effects in
this collision are negligible, the
results of our calculations can be compared with nonrelativistic data
obtained by
other authors (section \ref{subsec:H2}). The role of
the relativistic effects is investigated in sections \ref{subsec:Ne2},
\ref{subsec:Xe2}, and \ref{subsec:U2},
where the relativistic and nonrelativistic calculations of
the charge-transfer probabilities and cross sections are
performed for higher-$Z$ ions. 
%
\section{Theory}
\label{sec:theory}
%
\subsection{Two-center Dirac equation in the finite basis set}
\label{subsec:two-center}
%
\subsubsection{Two-center expansion}
Within the Born-Oppenheimer approximation,  the motion of the
electron is considered as a motion
in the field of the two nuclei being at given positions (the stationary case)
or moving along the classical trajectories (the non-stationary case).
Let $\vec{R}_A$ and  $\vec{R}_B$ are the positions of the target ($A$)
and projectile ($B$) nuclei, respectively. The time-dependent
$\Psi(\vec{r},t)$ and stationary $\psi(\vec{r})$ wave functions
are the solutions of the time-dependent and stationary Dirac equations,
respectively. In the atomic units ($\hbar=m=e=1$), these equations
are given by
\begin{equation}
i \frac{\partial \Psi(\vec{r},t)}{\partial t} = \hat h_{\rm D} \,
\Psi(\vec{r},t), \qquad
\hat h_{\rm D} \psi_n(\vec{r}) = \varepsilon_n \, \psi_n(\vec{r}) \,.
\end{equation}
Here $\varepsilon_n$ is the energy of the stationary state and
the $\hat h_{\rm D}$ is the two-center Dirac Hamiltonian defined by
\begin{equation}
\hat h_{\rm D} =c (\vec{\alpha} \cdot \vec{p}) \,+\, 
\beta \, c^2 \,+\, V_{AB}(\vec{r})\,,
\end{equation}
where $c$ is the speed of light, $\vec{\alpha}$, $\beta$ are the
Dirac matrices, and 
\begin{equation}
V_{AB}(\vec{r}) = V_{\rm nucl}^{A}(\vec{r}_A) + V_{\rm nucl}^{B}(\vec{r}_B)
\,, \qquad \vec{r}_A=\vec{r}- \vec{R}_A\,, \qquad \vec{r}_B=\vec{r}- \vec{R}_B,
\end{equation}
\begin{equation}
V_{\rm nucl}(\vec{r}) = \left  \{ 
\begin{array}{ll}  \displaystyle
-Z/r &  \quad \hbox{for the point nucleus}
\\[2mm] \displaystyle
\int d^3 \vec{r}^{\prime} \, 
\frac{\rho_{\rm nucl}(\vec{r}^{\prime})}{|\vec{r}-\vec{r}^{\prime}|} &
 \quad \hbox{for the extended nucleus. }
\end{array}
\right .
\end{equation}
The nuclear charge density $\rho_{\rm nucl}(\vec{r})$ is defined
by the nuclear model. 
In this paper we will use the Fermi model for the nuclear charge
distribution.

Here and in what follows we consider only the electric part of
the classical electromagnetic interaction between the electron and the moving
nuclei
neglecting the magnetic interaction ($e/c \, \vec{A}(\vec{r})$), which is
small for low-energy collisions.

The two-center expansion of the stationary wave function $\psi_n(\vec{r})$ and
the time-dependent wave function $\Psi(\vec{r},t)$ can be written in the form
\begin{equation}
\left \{
\begin{array}{lll} \displaystyle
\psi_n(\vec{r})  &=&  \displaystyle \sum_{\alpha=A,B} \, \sum_{a}
c^{n}_{\alpha a} \, \varphi_{\alpha,a} (\vec{r}-\vec{R}_\alpha)
\\[4mm]
\Psi(\vec{r},t)  &=&  \displaystyle \sum_{\alpha=A,B} \, \sum_{a}
C_{\alpha a}(t) \, \varphi_{\alpha,a} (\vec{r}-\vec{R}_\alpha(t)) \,,
\end{array}
\right .
\label{expan1}
\end{equation}
where index $\alpha=A,B$ enumerates the centers, index $a$
enumerates  basis functions at the given center, and
$\varphi_{\alpha,a} (\vec{r}-\vec{R}_\alpha)$ is the central-field
bispinor, centered at point $\alpha$.
The coefficients $c^{n}_{a \alpha}$ of the expansion (\ref{expan1}) 
for the stationary wave function $\psi_n(\vec{r})$ can be obtained
from the generalized eigenvalue equation
\begin{equation}
\sum_{k} H_{jk} \,c^{n}_{k} =  \displaystyle \varepsilon_n \,
\sum_{k} S_{jk} \,c^{n}_{k} \,,
\end{equation}
where indexes $j$ and $k$ enumerate the basis functions of both centers, and
the matrix elements of $H$ and $S$ are given by
\begin{equation}
H_{jk} \,=\, \langle j \mid \hat h_{\rm D} \mid k \rangle \,, \qquad
S_{j k} \,=\, \langle j \mid k \rangle \,.
\label{matr1}
\end{equation}
The expansion coefficients $C_{a \alpha}(t)$ of the time-dependent
wave function $\Psi(\vec{r},t)$ can be obtained by  solving
the linear system of first-order differential equations
\begin{equation}
i \sum_{k}S_{jk} \frac{dC_{k}(t)}{d t}  =  \sum_{k} \,
\left( H_{jk} - T_{jk}) \, C_{k}(t) \right ) \,.
\end{equation}
The matrix elements of $T$ are given by
\begin{equation}
T_{jk} \,=\, i \langle j \mid \frac{\partial}{\partial t} \mid k \rangle =
T^{\ast}_{kj} + i \frac{\partial}{\partial t} \, S_{jk} \,.
\label{matr2}
\end{equation}
Obviously the matrix $T$ is non-Hermitian, if 
the overlapping matrix $S$ depends on time.

The functions $\varphi_{\alpha}$ depend on time due to two reasons. 
First, the basis functions centered at the
target and projectile nuclei move together with
the nuclei. Second, the basis functions depend parametrically on
the distance between the nuclei, since their radial parts are obtained
from the radial equations, where for each center the potential of
the other nucleus is included in the so-called monopole approximation
(see section \ref{subsec:basis}). Therefore, the time derivative
of the basis function can be divided into two parts 
\begin{equation}
\Bigl \langle j \Bigl | \frac{\partial}{\partial t} \Bigr | k
\Bigr \rangle \,=\, \frac{dR}{dt} \,
\Bigl \langle \varphi_j \Bigl |  \frac{\partial \varphi_k}{\partial R}
\Bigr \rangle - \vec{v}_{\alpha_k}  \cdot 
\langle \varphi_j \mid \vec{\nabla} \mid \varphi_k \rangle \,,
\end{equation}
where  $\vec{v}_{\alpha} = d \vec{R}_{\alpha}/d t$ is the velocity of the ion
$\alpha$.
%
\subsubsection{Trajectories of nuclear motion}
\label{subsubsec:rutherford_sec}
\begin{figure}[hbt]
\centering
\includegraphics[width=10cm,clip]{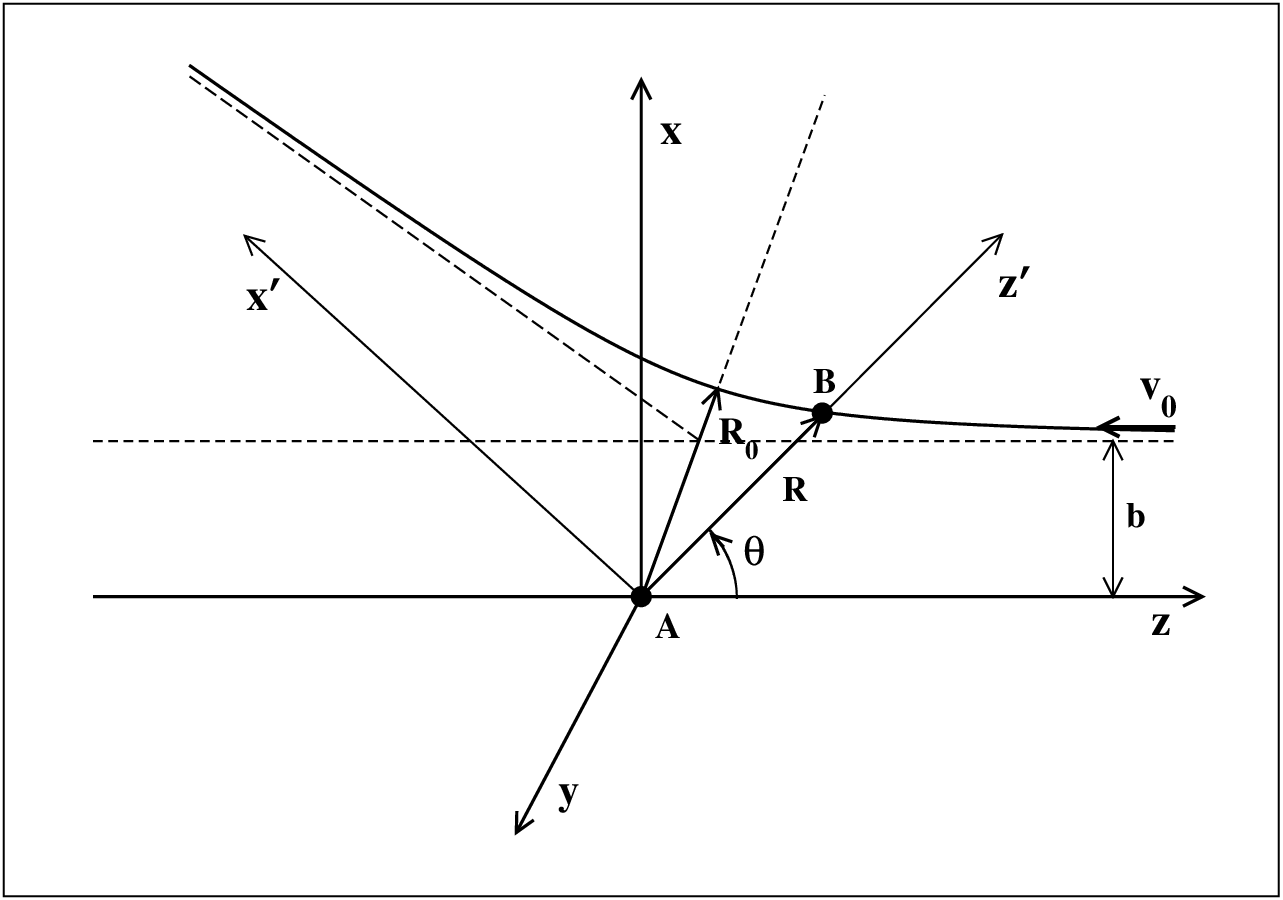}
\vspace{-2mm}
\caption{The hyperbolic Rutherford trajectory (b is the impact parameter,
R$_0$ is the minimal distance between target A and projectile B, v$_0$
is the initial projectile velocity). The coordinate system
$S_{\rm t}=(x,y,z)$ is defined with respect to the moving target ion.}
\label{rutherford}
\vspace{-2mm}
\end{figure}

In the ion-ion collisions the internuclear distance vector
$\vec{R}=\vec{R}_B-\vec{R}_A$, the length $R$ of vector $\vec{R}$,
the target velocity ($\vec{v}_A$), and the projectile velocity ($\vec{v}_B$)
are time dependent. This dependence is defined by the trajectories of
the nuclear motion. In low-energy collisions the nuclear trajectories
can be obtained by solving classical non-relativistic Newton's equations
of motion. In the case of point charges this solution is well-known
Rutherford hyperbola (see Fig.~\ref{rutherford}), which can be given
in the parametric representation  by the equations
\cite{Greiner_85}
%
%
\begin{equation}
\left \{
\begin{array}{lll} \displaystyle
R  &=&  \displaystyle a \, (\varepsilon \, \cosh \, \xi+1)
\\[2mm]
t &=&  \displaystyle \frac{a}{v}_{0} \, (\varepsilon \, \sinh \, \xi+\xi) \,,
\end{array}
\right. 
\label{rutherford1}
\end{equation}
where $\xi \in (-\infty,\infty)$,
\begin{equation}
a \,=\, \frac{Z_A \, Z_B \, e^2}{M_{\rm r} \, v^2_{0}} \,, \qquad
\varepsilon \,=\,\left (1+\frac{b^2}{a^2} \right)^{1/2},
\end{equation}
$v_0$ is the initial velocity of the projectile, $b$ is the impact
parameter,
and $M_{\rm r}$ is the reduced ion mass.
In the coordinate system $S_{\rm t}=(x,y,z)$, which is shown in
Fig.~\ref{rutherford}, the $X$ and $Z$ components of the internuclear distance
vector
$\vec{R}$ are given by
\begin{equation}
\left \{ \begin{array}{lll} \displaystyle
Z &=& \displaystyle R \, \cos \theta
\\[2mm] \displaystyle
X &=& \displaystyle R \, \sin \theta
\end{array} \right . \,, \qquad \hbox{where} \quad
\theta = 2 \, {\rm arctg}  \left[ \frac{ \sqrt{\varepsilon^2-1} \,
({\rm th}(\xi/2)+1)} {(\varepsilon+1)-(\varepsilon-1) \,
{\rm th}(\xi/2)} \right ] \,.
\end{equation}
The angle  $\theta$ is related to the scattering
angle $\Theta_{\infty}$ by
$\Theta_{\infty} = \pi - \theta(t=\infty)$.
\subsubsection{Time-dependent matrix Dirac equation}
%
In this work the two-center basis set $\varphi_j$ is not orthonormal.
Let us consider the transformation of the basis set $\varphi_j$
to the orthonormal basis $\varphi^{L}_j$ by a matrix $L^{-1}$
\begin{equation}
\varphi^{L}_j \,=\, \sum_{k} L^{-1}_{k j} \, \varphi_{k} \,, \qquad 
\varphi_j \,=\, \sum_{k} L_{k j} \, \varphi^{L}_{k} \,.
\end{equation}
Then the positive-defined matrix $S$ can be represented as the product
of $L^{+}$ and $L$:
\begin{equation}
S=L^{+} \, L \,, \qquad
S^{L}_{jk} = \langle \varphi^L_j \mid \varphi^{L}_k \rangle =
\bigl(L^{-1^{+}} S \, L^{-1} \bigr)_{jk} = \delta _{j,k} \,.
\label{Cholesky1}
\end{equation}
If the matrix $L$ is an upper-triangle matrix, then 
the decomposition (\ref{Cholesky1}) is so-called Cholesky
factorization \cite{Wilkinson_71}.
The expansion of the time-dependent wave function over the orthonormal
basis $\varphi^{L}_j$ is given by
\begin{equation}
\Psi(\vec{r},t)  =  \sum_{\alpha=A,B} \, \sum_{a}
C^{L}_{\alpha,a}(t) \,
\varphi^{L}_{\alpha,a} (\vec{r}-\vec{R}_\alpha(t),t) \,,
\label{expan2}
\end{equation}
where $\vec{C}^{L} = L\, \vec{C}$.

The time-dependent Dirac equation in the basis $\varphi^{L}_j$ can be
written in the form
\begin{equation}
i \, \frac{d \vec{C}^{L}(t)}{d t} \,=\, M \, \vec{C}^{L}(t) \,,
\label{dirac_time}
\end{equation}
where $ M=H^{L} - T^{L}$, the Hermitian Hamiltonian matrix $H^{L}$ is
\begin{equation} 
H^{L}_{ij} = \langle \varphi_{i} \mid \hat H \mid
\varphi_{j} \rangle = (L^{-1^{+}}\, H \, L^{-1})_{ij} \,,
\end{equation}
and  the matrix $T^{L}$  is defined by
\begin{equation} 
T^{L}_{ij} =
\langle \varphi^{L}_{i} \mid \hat T \mid
\varphi^{L}_{j} \rangle = (L^{-1^{+}}\, T \, L^{-1})_{ij} +
i \left (L \, \frac{dL^{-1}}{dt} \right)_{ij} =
\left( L^{-1^{+}} \left [ T  - i \, L^{^+} \frac{dL}{dt} \right] \,
L^{-1} \right)_{ij} \,.
\end{equation}
It should be noted that matrix $T^{L}$ is Hermitian, in contrast
to the matrix $T$ defined by Eq. (\ref{matr2}). Therefore, the matrix $M$
 is also Hermitian. 

The time-dependent matrix equation (\ref{dirac_time}) can be considered
as a linear system of the first-order differential equations at the range
of time $t \in (-\infty,\infty)$. 
We assume that at the initial moment of time ($t \to -\infty$) the electron
is localized on the target in the $1s$ state and the projectile is the bare
nucleus. Then, the wave function $\Psi(\vec{r},t)$  
at $t \to -\infty$
is given by
\begin{equation}
\Psi(\vec{r},t) \Bigr |_{t \to -\infty} = \psi_{1s}(\vec{r}).
\end{equation}
If the Dirac $1s$-target wave function $\psi_{1s}(\vec{r})$ is
included in the basis set, the initial conditions for the expansion
coefficients can be written as
\begin{equation}
C^{L}_j(t) \Bigr |_{t \to -\infty} = C_j(t) \Bigr |_{t \to -\infty} =
\delta_{j,1s} \,.
\end{equation}
Equation (\ref{dirac_time}) is solved numerically,
using  approximate evolution operator
\begin{equation} 
\vec{C}^{L}(t+\Delta t) = e^{-i \overline M \, t} \,
\vec{C}^{L}(t) + O(\Delta^3 t) \,,
\end{equation}
where Hermitian matrix $\overline M$ is chosen as
\begin{equation} 
\overline M = M(t+\Delta t/2).
\end{equation}
Since the approximate evolution operator $U(t)= \exp{(-i \overline M \, t)}$
is unitary, the time-dependent wave function conserves the
norm at each time step
\begin{equation} 
\langle \Psi(\vec{r},t)|  \Psi(\vec{r},t) \rangle =
\sum_{j} |C^{L}_j(t)|^2=1 \,.
\end{equation}
The matrix $ e^{-i\overline M t} $ is calculated at each time step using
the eigen decomposition of matrix $\overline M$
\begin{equation}
\overline M \,=\, V \, \Lambda \, V^{+} \,,
\end{equation}
where $\Lambda$ is a diagonal matrix and columns of matrix $V$ are
the eigenvectors of $\overline M$. Then one obtains
\begin{equation}
e^{-i\overline M t} \,=\, V \, e^{-i \Lambda t} \, V^{+} \,.
\end{equation}
The time grid points $t_i$ are chosen as
$t_i~=~a/v_0~(\varepsilon~\sinh~\xi_i~+~\xi_i)$, where the parameter
$\xi$ runs a uniform grid. The grid points $R_i$
can be obtained using equation (\ref{rutherford1}).
%
\subsection{Basis functions}
\label{subsec:basis}
In our approach the basis set contains Dirac and Dirac-Sturm orbitals.
The Dirac-Sturm orbitals can be considered as pseudo-states, which should
be included in the basis to take into account the contribution of the
positive- and negative-energy Dirac continuum. Both types of basis
functions $\varphi_{\alpha a}$ are the central field Dirac bispinors
centered at the position $\vec{R}_{\alpha}$ ($\alpha=A,B$)
\begin{equation}
\varphi_{n\kappa m}(\vec{r}) =
\left  ( \begin{array}{l} \displaystyle
\,\, \frac{~P_{n \kappa}(r)}{r} \,  \chi_{\kappa m}(\Omega,\sigma)
\\[4mm] \displaystyle
i \, \frac{Q_{n \kappa}(r)}{r} \, \chi_{-\kappa m}(\Omega,\sigma)
\end{array} \right ) \,,
\end{equation}
where $P_{n \kappa}(r)$ and $Q_{n \kappa}(r)$ are large and small radial
components, respectively, and $\kappa=(-1)^{l+j+1/2}(j+1/2)$ is the
relativistic angular quantum number.
The large and small radial orbital components are obtained by solving
numerically the Dirac or Dirac-Sturm equations in the central field
potential $V(r)$. The Dirac equation is given by 
\begin{equation}
\left  \{ 
\begin{array}{lll}  \displaystyle
c \left (-\frac{d}{dr}+\frac{\kappa}{r} \right ) \,Q_{n\kappa}(r) +
\left (V(r)+c^2 \right ) \, P_{n\kappa}(r) &=&
\varepsilon_{n \kappa} \, P_{n \kappa}(r)
\\[4mm] \displaystyle
c \left(~~\frac{d}{dr}+\frac{\kappa}{r} \right) \, P_{n \kappa}(r) +
\left (V(r)-c^2 \right ) \, Q_{n\kappa}(r) &=&
\varepsilon_{n \kappa} \, Q_{n \kappa}(r)\,. \end{array} \right  .
\label{dirac1}
\end{equation}
The radial components of the Dirac-Sturm orbitals which we denote by
$\overline P_{n\kappa}(r)$ and  $\overline Q_{n\kappa}(r)$ are the
solutions of the Dirac-Sturm generalized eigenvalue equation
\begin{equation}
\left  \{
\begin{array}{lll}  \displaystyle
c \left (-\frac{d}{dr}+\frac{\kappa}{r} \right ) \,\overline Q_{n\kappa}(r) +
\left (V(r)+c^2 - \varepsilon_{n_0\kappa} \right) \,
\overline P_{n\kappa}(r) &=&
\lambda_{n \kappa} \, W_{\kappa}(r) \, \overline P_{n \kappa}(r)
\\[3mm] \displaystyle
c \left(~~\frac{d}{dr}+\frac{\kappa}{r} \right) \, \overline P_{n \kappa}(r) +
\left (V(r)-c^2  - \varepsilon_{n_0 \kappa} \right) \,
\overline Q_{n \kappa}(r) &=&
\lambda_{n \kappa} \, W_{\kappa}(r) \, \overline Q_{n \kappa}(r)\,.
\end{array}
\right .
\label{sturm1}
\end{equation}
Here $\lambda_{n \kappa}$ can be considered as the eigenvalue of the
Dirac-Sturm operator and $W_{\kappa}(r)$ is a constant sign weight function.
The energy $\varepsilon_{n_0 \kappa}$ is fixed in the Dirac-Sturm equation.
 If  $W(r) \to 0$ at $r \to \infty$, all Sturmian functions have the
same asymptotic at $r \to \infty$. It is clear that for
$\lambda_{n \kappa}=0$ the Sturmian function coincides with the reference
Dirac orbital which has the radial parts $P_{n_0 \kappa}(r)$ and $Q_{n_0
\kappa}(r)$.
The widely known choice of the weight function is $W(r)=1/r$, which leads
to the well known 'charge quantization'
$Z^{\ast}_{n \kappa}=Z+\lambda_{n\kappa}$. The main advantage of this choice
for the Coulomb potential $V(r)=-Z/r$ is that
the Coulomb-Sturmian orbitals can be given in an analytical form. 
This is not the case, however, for the non-Coulomb
potential $V(r)$. In the relativistic case the choice $W(r)=1/r$ is not very
successful, because of the
incorrect behavior of the Coulomb-Sturmian orbitals at $r \to 0$. 
For this reason the standard form of the equation has to be
modified
\cite{Drake_88, Grant_91, Szmytkowski_97}.

In our calculations we use the following weight function
\begin{equation}
W_{\kappa}(r)  \,=\, - \, \frac{1 \,-\, 
\exp(-(\alpha_{\kappa} \, r)^2)}{(\alpha_{\kappa} \, r)^2}\,.
\label{sturm2}
\end{equation}
In contrast to $1/r$, this weight function is regular at origin.
It is well-known that the Sturmian operator is Hermitian and does
not contain continuum spectra, in contrast to the Dirac operator.
Therefore, the set of the Sturmian eigenfunctions forms the discrete
and complete basis set of one-electron wave functions.

The central-field potential $V(r)$ in equations (\ref{dirac1}) and
(\ref{sturm1}) is arbitrary, and, therefore, it can be chosen 
to provide most appropriate Dirac and Dirac-Sturm basis orbitals. 
At short internuclear distances the wave function of the electron
experiences the strong Coulomb field of both nuclei.
To take into account this effect we 
have included the Coulomb potential of the second ion
in the total one center potential $V(r)$ in so-called monopole
approximation. For instance, the total central-field potential
$V^{A}(r)$ of the center $A$ is given by
\begin{equation}
V^{A}(r) = V^{A}_{\rm nucl}(r) + V^{B}_{\rm mon}(r) \,,
\end{equation}
where $V^{A}_{\rm nucl}(r)$ is the Coulomb potential of the nucleus $A$ and
$V^{B}_{\rm mon}(r)$ is the spherically-symmetric part of the
reexpansion  of the potential
$V^{B}_{\rm nucl}(\vec{r}-\vec{R_B})$ on the center $A$
\begin{equation}
V^{B}_{\rm mon}(r) = \frac{1}{4 \pi} \,\int d\Omega_A \,\,
V^{B}_{\rm nucl}(\vec{r}-\vec{R_B}) \,. 
\end{equation}
For the point nucleus the potential $V^{B}_{\rm mon}(r)$ is given by
\begin{equation}
V^{B}_{\rm mon}(r) =
\left \{
\begin{array}{ll} \displaystyle \,
-\frac{Z_B}{r} \quad & r \ge R
\\[4mm] \displaystyle
-\frac{Z_B}{R} \quad & r< R \,.
\end{array} \right .
\end{equation}
%
\subsection{Two-center integrals}
\label{subsec:integrals}
The matrix elements of $H$ and $S$
(Eq. \ref{matr1}) are easily reduced to radial integrals \cite{Grant_70},
which are calculated  by numerical integration in the radial
semi-logarithmic grid \cite {Bratsev_77}.
%
\subsubsection*{Modified L\"owdin reexpansion procedure}
\begin{figure}[hbt]
\centering
\includegraphics[width=10cm,clip]{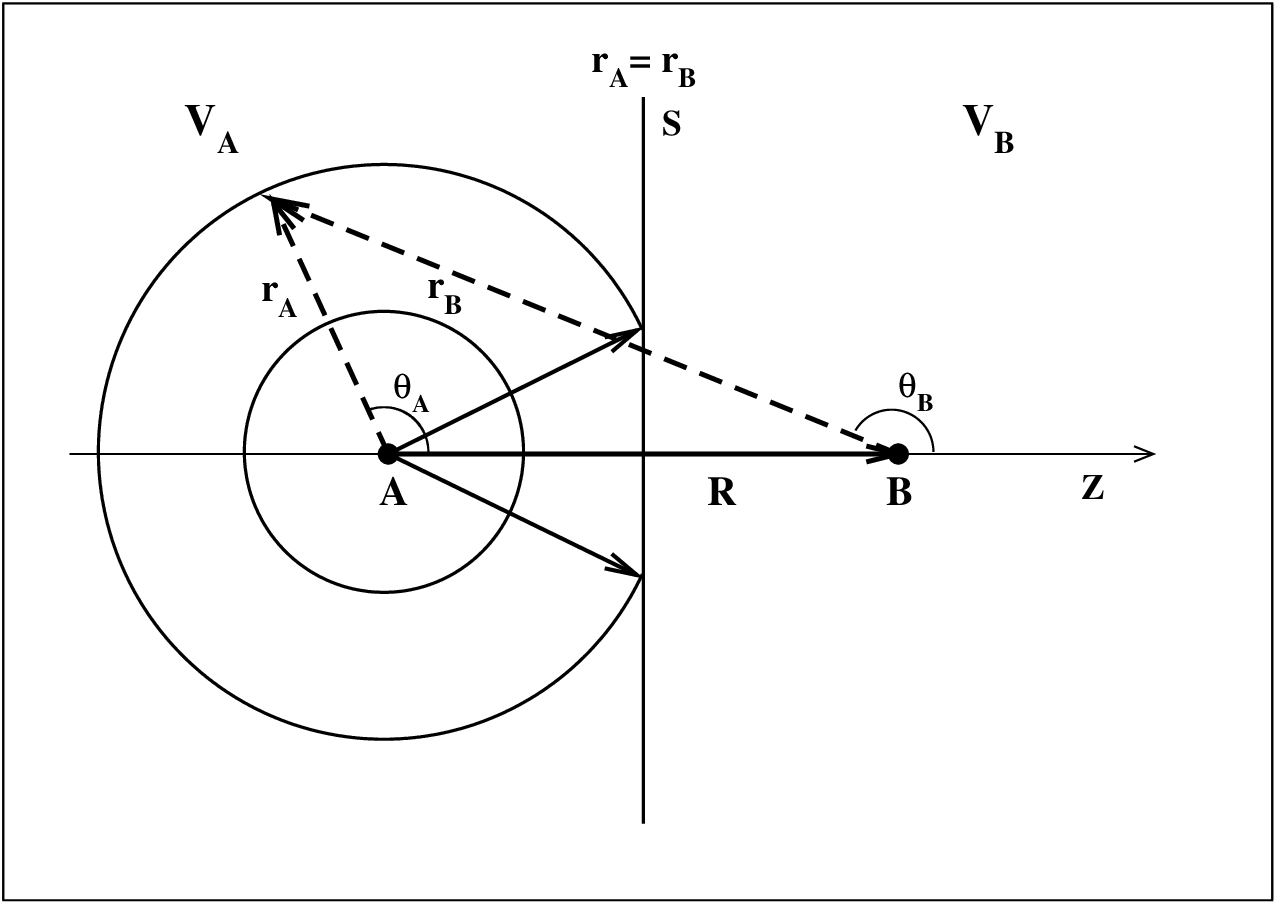}
\caption{Integration regions $V_A$ and $V_B$.}
\label{int_region}
\vspace{-2mm}
\end{figure}

Two-center matrix elements are calculated using a symmetrical
reexpansion procedure, proposed in Refs. \cite{Kotochigova_95, Tupitsyn_98}.
The reexpansion procedure is based on the technique developed by L\"owdin
\cite{Lowdin_56}. We assume that in the local coordinate frame
the $z$-axis is directed along the internuclear axis $A$--$B$
(see Fig. \ref{int_region}). The following geometrical relations take
place
\begin{equation}
\begin{array}{lll} \displaystyle
\vec{r}_A = \vec{r} - \vec{R}_A \,, \qquad
\vec{r}_B = \vec{r} - \vec{R}_B \,, \qquad
\vec{R} = \vec{R}_B - \vec{R}_A \,,
\\[3mm] \displaystyle
\cos \theta_A= \frac{r_A^2+R^2-r^2_B}{2Rr_A} \,, \qquad
\cos \theta_B= \frac{r_A^2-R^2-r^2_B}{2Rr_B} \,.
\end{array}
\end{equation}
%
%
Let the indexes $a$ and $b$ enumerate basis functions centered at the
points $A$ and $B$, respectively. The standard L\"owdin reexpansion of the
nonrelativistic central-field function $F_b(\vec{r}_B)$ centered at
the point $B$ in terms of the spherical harmonics $Y_{lm}(\vec{r}_A)$ centered 
at the point $A$ can be written in the form \cite{Lowdin_56, Sharma_76}
\begin{equation}
F_b(\vec{r}_B) = \frac{f_b(r_B)}{r_B} \, Y_{l_bm_b}(\theta_B, \varphi) =
\frac{1}{r_A} \, \sum_{l=0}^{\infty} \, \alpha_{lm_b}(f_b,l_b|r_A) \,
Y_{l m_b}(\theta_A, \varphi) \,,
\end{equation}
where $\alpha_{lm_b}(f_b,l_b|r_A)$ is so-called L\"owdin $\alpha$-function
defined by
\small
\begin{equation}
\alpha_{lm_b}(f_b,l_b|r_A) = \frac{K_{l_bm_b}  K_{lm_b}}{R}
\int \limits_{|r_A-R|}^{|r_A+R|} f_b(r) \,
P^{|m_b|}_{l_b} \left(\frac{r_A^2-R^2-r^2}{2r R} \right) \,
P^{|m_b|}_l\left (\frac{r_A^2+R^2-r^2}{2r_AR} \right) dr \,.
\label{alpha1}
\end{equation}
\normalsize
Here $P_{l}^{|m|}$ is the standard associated Legendre polynomial and
$K_{lm}$ is the normalization constant
\begin{equation}
\nonumber
K_{lm}=\sqrt{\frac{2l+1}{2}\frac{(l-|m|)!\!}{(l+|m|)!\!}} \,.
\end{equation}
Similarly, the function $F_a({\vec r}_A)$ centered at point $A$ can be
expanded in spherical harmonics $Y_{lm}(\theta_B, \varphi)$ centered 
at the point $B$.

When the logarithmic or semi-logarithmic grid is used the radial grid
step increases with increasing radius $r_A$. Therefore, the L\"owdin
reexpansion procedure becomes unstable and poorly convergent for the
values of radius $r_A$ in the region near $r_A=R$, especially for
the oscillating and strongly localized atomic-like wave functions.
In addition, the L\"owdin procedure is not symmetric
with respect to the centers $A$ and $B$.

To improve the convergence we modified the standard
L\"owdin reexpansion procedure by dividing the range of the integration into
two regions $V_A$ and $V_B$ as shown in  Fig. \ref{int_region}.
The region $V_A$ contains ion $A$ and the region $V_B$ contains ion $B$.
The dividing of the integration area into two parts can be done, for example,
by a plane passing through the center of the segment ($AB$). We apply
the reexpansion procedure only to the "tails" of the wave functions occurring
in a given region. To describe this procedure we introduce the step-wise
functions $\Theta_A(\vec{r})$ and $\Theta_B(\vec{r})$ by
  \begin{equation}
  \Theta_A(\vec{r}) =\left \{
  \begin{array}{c}
  1 \quad \vec{r} \in V_A \\
  0 \quad \vec{r} \in V_B\,,
  \end{array} \right.
  \qquad \Theta_B(\vec{r}) =\left \{
  \begin{array}{c}
  0 \quad \vec{r} \in V_A \\
  1 \quad \vec{r} \in V_B\,,
  \end{array} \right.
  \end{equation}
and rewrite the product of the functions centered at the different points
 in the following way
\begin{equation}
F_a(\vec{r}_A) \cdot F_b(\vec{r}_B) = F_a(\vec{r}_A) \cdot
(F_b(\vec{r}_B) \, \Theta_A(\vec{r})) + 
(F_a(\vec{r}_A) \,\Theta_B(\vec{r})) \cdot  F_b(\vec{r}_B) \,.
\end{equation}
The reexpansion of the function tail $F_b(\vec{r}_B) \, \Theta_A(\vec{r})$
centered at $B$ onto center $A$ has the form
\begin{equation}
F_b(\vec{r}_B) \, \Theta_A(\vec{r}) =
\frac{f_{b}(r_B)}{r_B} \, Y_{l_b m_b}(\theta_{B}, \varphi) \, 
\Theta_A (\vec{r}) = \frac{1}{r_{A}} \sum_{l}
\overline{\alpha}_{lm_b}(f_b,l_b|r_{A}) \cdot Y_{l m_b}(\theta_{A}, \varphi),
\label{lowdin1}
\end{equation}
where
\small
\begin{equation}
\overline \alpha_{lm_b}(f_b,l_b|r_A) = \frac{K_{l_bm_b}  K_{lm_b}}{R}
\int \limits_{r^A_{>}}^{|r_A+R|} f_b(r) \,
P^{|m_b|}_{l_b} \left(\frac{r_A^2-R^2-r^2}{2r R} \right) \,
P^{|m_b|}_l\left (\frac{r_A^2+R^2-r^2}{2r_AR} \right) dr \,
\label{alpha2}
\end{equation}
\normalsize
and $r^{A}_{>} = \max\{r_A, |r_A-R|\}$.

In the relativistic case the spin-angular part $\chi_{\kappa m}$
of the large and small components of the central-field wave function
is the Pauli spinor \cite{Rose_61}
\begin{equation}
\chi_{\kappa \mu}(\vec{r},\sigma) =\chi_{lj \mu}(\vec{r},\sigma) =
\sum_{m,m_s} C^{j\mu}_{lm, \frac{1}{2},m_s} \, Y_{lm}(\vec{r}) \,
\Phi_{m_s}(\sigma) \,,
\end{equation}
where $C^{j\mu}_{lm, \frac{1}{2},m_s}$ are the Clebsch-Gordan coefficients
\cite{Varshalovich_88} and $\Phi_{m_s}(\sigma)$ is a spin function.

The symmetric reexpansion of the relativistic wave function
``tails'' onto centers $A$ and $B$ can be written in the form
\small
\begin{equation}
\left  ( \begin{array}{l} \displaystyle
\,\,\frac{P_{b}(r_B)}{r_B} \, \chi_{\kappa_b \mu_b}(\vec{r}_{B})
\\[4mm] \displaystyle
i \,\frac{Q_{b}(r_B)}{r_B} \, \chi_{-\kappa_b \mu_b}(\vec{r}_{B}) \, 
\end{array} \right ) \,\Theta_A (\vec{r}) =
\sum_{\kappa}
\left  ( \begin{array}{l} \displaystyle
\,\, \frac{\overline p_{\kappa \mu_b}(b|r_A)}{r_A} \, 
\chi_{\kappa \mu_b}(\vec{r}_{A})
\\[4mm] \displaystyle
i \,\frac{\overline q_{-\kappa \mu_b}(b|r_A)}{r_A} \,
\chi_{-\kappa \mu_b}(\vec{r}_{A}) \, 
\end{array} \right ) \,
\,
\end{equation}
\label{rel_lowdin1}
\normalsize
and
\small
\begin{equation}
\left  ( \begin{array}{l} \displaystyle
\,\,\frac{P_{a}(r_A)}{r_A} \, \chi_{\kappa_a \mu_a}(\vec{r}_{A})
\\[4mm] \displaystyle
i \,\frac{Q_{a}(r_A)}{r_A} \, \chi_{-\kappa_a \mu_a}(\vec{r}_{A}) \, 
\end{array} \right ) \,\Theta_B (\vec{r}) =
\sum_{\kappa} (-1)^{l_a-l}
\left  ( \begin{array}{l} \displaystyle
\,\, \frac{\overline p_{\kappa \mu_a}(a|r_B)}{r_B} \, 
\chi_{\kappa \mu_a}(\vec{r}_{B})
\\[4mm] \displaystyle
i \,\frac{\overline q_{-\kappa \mu_a}(a|r_B)}{r_B} \,
\chi_{-\kappa \mu_a}(\vec{r}_{B}) \, 
\end{array} \right ) \,.
\end{equation}
\label{rel_lowdin2}
\normalsize
The $\overline p$- and  $\overline q$-functions, which are the relativistic
analogs of the modified L\"owdin $\overline \alpha$-functions, are defined
by
\begin{equation}
\left \{
\begin{array}{lll} \displaystyle
\overline p_{\kappa \mu_b}(b|r_{A}) &=& \displaystyle
\sum_{m_b, m_s}  C^{j_b \mu_b}_{l_bm_b,\frac{1}{2}m_s} \,
C^{j \mu_b}_{lm_b,\frac{1}{2} m_s} \, \overline{\alpha}_{lm_b}(P_b, l_b |r_{A})
\\[4mm]  \displaystyle
\overline q_{ \kappa \mu_b}(b|r_{A})
&=& \displaystyle
\sum_{m_b, m_s} C^{j_b \mu_b}_{\overline l_b m_b,\frac{1}{2}m_s} \,
C^{j \mu_b}_{\overline l m_b,\frac{1}{2} m_s} \,
\overline{\alpha}_{\overline l m_b}(Q_b \, \overline l_b |r_{A}) \,,
\label{pq-functions}
\end{array} \right .
\end{equation}
where $\overline l = l-{\rm sign}(\kappa)$. Functions
$\overline{\alpha}_{lm_b}(P_b, l_b |r_{A})$ and
$\overline \alpha_{\overline l m_b}(Q_b, \overline l_b |r_{A})$
are defined by equation (\ref{alpha2}), where the function
$f_b(r)$ has to be replaced by the functions $P_b(r)$ and  $Q_b(r)$, respectively. 
Functions $\overline p_{ \kappa \mu_a}(a|r_{B})$ and
$\overline q_{ \kappa \mu_a}(a|r_{B})$ are defined similarly to equation
(\ref{pq-functions}), where indices $A$ and $b$ should be replaced by
$B$ and $a$, respectively.
%
\subsubsection*{Two-center overlap integrals}
Let us consider two-center overlap integrals $S^{(0)}_{ab}$.
Here and below symbol (0)  means that the integral is considered in the
local coordinate frame, where the $z$-axis is directed along internuclear
axis $A$--$B$.

The integral $S^{(0)}_{ab}$ can be divided into two parts
\begin{equation}
S^{(0)}_{ab} = \langle a \mid \ b \rangle \,=\,\langle a \mid \ b \rangle_{A}
\,+\, \langle a \mid \ b \rangle_{B} \,,
\end{equation}
where the notations $<>_A$ and  $<>_B$ mean the integration over the
regions $V_A$ and $V_B$, respectively (see Fig. \ref{int_region}).
Using the reexpansions of the large
and small components onto the center $A$ (in the region $V_A$) and
onto the center $B$ (in the region $V_B$) we obtain
\begin{equation}
\left \{
\begin{array}{llll} \displaystyle
\langle a \mid b \rangle_{A} &=& &\displaystyle \delta_{\mu_a,\mu_b} 
\int \limits_{0}^{\infty} dr \, \left[
P_a(r) \cdot \overline p_{\kappa_a \mu_a}(b|r) +  Q_a(r) \cdot
\overline q_{\kappa_a \mu_a}(b|r) \right ]
\\[4mm] \displaystyle
\langle a \mid b \rangle_{B} &=&  \displaystyle (-1)^{l_b-l_a}
&  \displaystyle \delta_{\mu_a,\mu_b} \, \displaystyle
\int \limits_{0}^{\infty} dr \, \left[ 
P_b(r) \cdot \overline p_{\kappa_b \mu_a}(a|r) +
Q_b(r) \cdot \overline q_{\kappa_b \mu_a}(a|r)
\right ]\,.
\end{array} \right .
\end{equation}
Matrix elements of the nuclear attraction potentials,
$V^{A}_{\rm nucl}(r_A)$ and
$V^{B}_{\rm nucl}(r_B)$, and of the
mass operator $\beta mc^2$are calculated similarly to the
overlap integral.
\subsubsection*{Two-center gradient matrix elements}
%
As in case of the overlap integral, the region of integration
for the gradient matrix element $G^{(0)}_{ab}(q)$ is divided into two parts,
\begin{equation}
G^{(0)}_{ab}(q)=\langle a \mid \vec{\nabla}_q \mid \ b \rangle =
\langle a \mid \vec{\nabla}_q \mid b \rangle_{A} +
\langle a \mid \vec{\nabla}_q \mid b \rangle_{B}\,.
\end{equation}
Here index $q=1,0,-1$ enumerates covariant spherical coordinates.
Using the Gauss theorem \cite{Marsden_03} for the integration over region
$A$, we obtain
\begin{equation}
G^{(0)}_{ab}(q)= - \langle b \mid \vec{\nabla}_q \mid a \rangle_{A} + 
\langle a \mid \vec{\nabla}_q \mid b \rangle_{B} +
\delta_{q,0} \, \langle a \mid b \rangle_{S} \,,  
\label{nabla1}
\end{equation}
where $\langle a \mid b \rangle_{S}$ is the surface integral over
the region $S$ (see Fig.~\ref{int_region}).
The volume integrals over regions $A$ and $B$ are given by
\small
\begin{equation}
\begin{array}{lll} \displaystyle
\langle b \mid \vec{\nabla}_{q} \mid a \rangle_{A} &=&  \displaystyle 
\sumpr_{\kappa}  g^{1 q}(j \mu_b,j_a \mu_a)
\int \limits_{0}^{\infty} dr \Bigl [
\overline p_{\kappa \mu_b}(b|r) \hat D_{\kappa,\kappa_a} P_a(r) +
\overline q_{\kappa \mu_b}(b|r) \hat D_{-\kappa,-\kappa_a} Q_a(r) \Bigr] \,,
\\[3mm] \displaystyle
\langle a \mid \vec{\nabla}_{q} \mid b \rangle_{B} &=&  \displaystyle 
\sumpr_{\kappa} (-1)^{l_a-l} g^{1 q}(j \mu_a,j_b \mu_b)
\int \limits_{0}^{\infty} dr \Bigl [
\overline p_{\kappa \mu_a}(a|r) \hat D_{\kappa,\kappa_b} P_b(r) +
\overline q_{\kappa \mu_a}(a|r) \hat D_{-\kappa,-\kappa_b} Q_b(r) \Bigr],
\end{array}
\label{nabla2}
\end{equation}
\normalsize
where the prime at the sum symbol indicates
that the summation is restricted to odd  values of $l_a+l$ and $l_b+l$,
and the operator $D_{\kappa,\kappa^{\prime}}$ 
is defined by
\begin{equation}
D_{\kappa,\kappa^{\prime}} = \frac{d}{d r} + 
\frac{\kappa^{\prime}(\kappa^{\prime}+1)-\kappa(\kappa+1)}{2r} \,.
\end{equation}
The coefficients $g^{kq}(j \mu,j^{\prime} \mu^{\prime})$ are the
relativistic analogs of the Gaunt coefficients \cite{Condon_35}
\begin{equation}
g^{kq}(j \mu,j^{\prime} \mu^{\prime}) = 
\frac{\sqrt{(2j+1)(2j^{\prime}+1)}}{2k+1} \, (-1)^{\frac{1}{2}+\mu^{\prime}} \,
C^{k0}_{j -\frac{1}{2}, j^{\prime} \frac{1}{2}} \,
C^{kq}_{j \mu,j^{\prime} -\mu^{\prime} }.
\end{equation}
The relativistic Gaunt coefficient is non-zero only if $l+l^{\prime}+k$ is even.

The surface integral is given by
\begin{equation}
\begin{array}{ll} \displaystyle
\langle a \mid b \rangle_{S}  =  \delta_{\mu_a,\mu_b} & \displaystyle 
\frac{1}{2} \Bigl [
\sum_{m,m_s} C^{j_a\mu_a}_{l_a m, \frac{1}{2} m_s} \,
C^{j_b\mu_a}_{l_b m, \frac{1}{2} m_s}
\int \limits_{R/2}^{\infty} dr \, \frac{1}{r} \, P_a(r) \,
P_b(r) \, U_{l_al_b m} \Bigl( \frac{R}{2r} \Bigr)
\\[5mm] & \times \displaystyle
\sum_{m,m_s} C^{j_a\mu_a}_{\overline l_a m, \frac{1}{2} m_s} \, 
C^{j_b\mu_a}_{\overline l_b m, \frac{1}{2} m_s}
\int \limits_{R/2}^{\infty} dr \, \frac{1}{r} \, Q_a(r) \,
Q_b(r) \, U_{\overline l_a \overline l_b m} \Bigl( \frac{R}{2r}\Bigr)
\Bigr ] \,,
\end{array}
\end{equation}
where
\begin{equation}
U_{l_a,l_b,m}(x) \,=\, (-1)^{l_b-m} \,
\sqrt{(2l_a+1)(2 l_b+1)} \, K_{l_a|m|} K_{l_b|m|} \,
P^{|m|}_{l_a}(x) \, P^{|m|}_{l_b}(x) \,.
\end{equation}
%
\subsubsection*{Two-center ($\vec{\alpha}\cdot\vec{p}$) matrix elements}
Two-center ($\vec{\alpha}\cdot\vec{p}$) matrix elements $A^{(0)}_{ab}$ can
be divided into three parts, similarly to the gradient matrix elements
$G^{(0)}_{ab}(q)$ (\ref{nabla1}),
\begin{equation}
A^{(0)}_{ab} = \langle a \mid \vec{\alpha}\cdot\vec{p} \mid \ b \rangle =
\langle b \mid \vec{\alpha} \cdot\vec{p} \mid a \rangle_{A} \,+\, 
\langle a \mid \vec{\alpha} \cdot\vec{p} \mid b \rangle_{B} \,+\,
\frac{1}{i} \, \langle a \mid \vec{\alpha}_0 \mid b \rangle_{S} \,,  
\label{alp1}
\end{equation}
where the volume integrals
$\langle b \mid \vec{\alpha}\cdot \vec{p} \mid a \rangle_{A}$ 
and $\langle b \mid \vec{\alpha} \cdot\vec{p} \mid a \rangle_{A}$ are given by
\begin{equation}
\begin{array}{llr} \displaystyle
\langle b \mid \vec{\alpha}\cdot \vec{p} \mid a \rangle_{A} &=& \displaystyle
\delta_{\mu_a,\mu_b}  \int \limits_{0}^{\infty} dr
\Bigl[ \Bigl( -\frac{dQ_a}{d r} + \frac{\kappa_a Q_a}{r} \Bigr)
\overline p_{\kappa_a \mu_a}(b|r) +
\Bigl( \frac{dP_a}{d r} + \frac{\kappa_a P_a}{r} \Bigr )
\overline q_{\kappa_a \mu_a}(b |r) \Bigr] \,,
\\[3mm] \displaystyle
\langle a \mid \vec{\alpha} \cdot\vec{p} \mid b \rangle_{B} &=& \displaystyle
(-1)^{l_a-l_b} \delta_{\mu_a,\mu_b}
\int \limits_{0}^{\infty} dr
\Bigl[ \Bigl( -\frac{dQ_b}{d r} + \frac{\kappa_b Q_b}{r} \Bigr )
\overline p_{\kappa_b \mu_a}(a|r) +
\Bigl( \frac{dP_b}{d r} + \frac{\kappa_bP_b}{r} \Bigr)
\overline q_{\kappa_b \mu_a}(a |r) \Bigr]\,.
\end{array}
\end{equation}
The last term in equation (\ref{alp1}) is the surface integral, 
which is given by
\begin{equation}
\begin{array}{lll} \displaystyle
\frac{1}{i} \, \langle a \mid \vec{\alpha}_0 \mid b \rangle_{S}
&=&  \displaystyle 
\delta_{\mu_a,\mu_b} \sum_{m,m_s} m_s 
C^{j_a\mu_a}_{\overline l_a m, \frac{1}{2} m_s} \,
C^{j_b\mu_a}_{l_b m, \frac{1}{2} m_s}
\int \limits_{R/2}^{\infty} dr \, \frac{1}{r} \, Q_a(r) \, P_b(r) \,
U_{\overline l_a l_b m}  \Bigl( R/2r \Bigr )
\\[5mm] &+&  \displaystyle
\delta_{\mu_a,\mu_b} \, \sum_{m,m_s} m_s 
C^{j_a\mu_a}_{l_a m, \frac{1}{2} m_s} \,
C^{j_b\mu_a}_{\overline l_b m, \frac{1}{2} m_s}
\int \limits_{R/2}^{\infty} dr \, \frac{1}{r} \, P_a(r) \, Q_b(r) \,
U_{l_a \overline l_b m} \Bigl( R/2r \Bigr ) \,.
\end{array}
\end{equation}
%
\subsubsection*{Transformation of the two-center matrix elements to the
laboratory frame}
%
As indicated above, the laboratory frame  $S$
is defined to be at rest with respect to the initial target position. Then,
the two-center matrix elements calculated in the local coordinate frame 
$S^{\prime}=(x^{\prime},y^{\prime},z^{\prime})$ (see Fig.~\ref{rutherford})
have to be transformed to the laboratory frame $S=(x,y,z)$.
The corresponding two-center integrals can be obtained from
 $S^{(0)}_{ab}$, $H^{(0)}_{ab}$, and $G^{(0)}_{ab}$
by rotating the coordinate system 
around the y-axis
for angle $-\theta$ (Fig.~\ref{rutherford}).
For the overlap integrals $S_{ab}$ and the two-center Dirac-Hamiltonian matrix
elements $H_{ab}$ in the laboratory frame $S$ we obtain
\begin{equation}
\begin{array}{lll} \displaystyle
S_{a,b} &=& S_{n_a \kappa_a \mu_a; \, n_b \kappa_b \mu_b} = 
 \displaystyle \sum_{\mu}
d^{j_a}_{\mu_a \mu}(\theta) \,\, d^{j_b^{\ast}}_{\mu_b \mu}(\theta) \,
S^{(0)}_{n_a\kappa_a \mu; \, n_b \kappa_b \mu} \,,
\\[3mm] \displaystyle
H_{a,b} &=& H_{n_a \kappa_a \mu_a; \, n_b \kappa_b \mu_b} =
\displaystyle \sum_{\mu}
d^{j_a}_{\mu_a \mu}(\theta) \,\, d^{j_b^{\ast}}_{\mu_b \mu}(\theta) \,
H^{(0)}_{n_a\kappa_a \mu; \, n_b \kappa_b \mu} \,,
\end{array}
\end{equation}
where $d^{j}_{\mu^{\prime} \mu}(\theta)$ are real Wigner's D-functions
\cite{Varshalovich_88}.
The transformation of the gradient matrix elements $G^{(0)}_{ab}(q)$
to the laboratory frame $S$ is given by
\begin{equation}
G_{ab}(q) = G_{n_a \kappa_a \mu_a; \, n_b \kappa_b \mu_b}(q) =
\sum_{\mu_a^{\prime},\mu_b^{\prime},q}
d^{j_a}_{\mu_a \mu_a^{\prime}}(\theta) \,\, 
d^{j_b}_{\mu_b \mu_b^{\prime}}(\theta) \,
d^{1}_{q{\prime}q}(\theta) \,
G^{(0)}_{n_a \kappa_a \mu_a^{\prime}; \, n_b \kappa_b \mu_b^{\prime}}(q) \,.
\end{equation}
%
%
\subsection{Charge-transfer probabilities and cross sections}
\label{subsec:prob}
\subsubsection{Transition amplitudes}
%
Transition amplitude for electron capture to an ion state $\alpha n$
is given by
\begin{equation}
T_{\alpha n}(t) = \langle \psi_{\alpha n}(\vec{r},t) \mid
\Psi(\vec{r},t) \rangle \,, \qquad t \to \infty \,.
\label{amplit1}
\end{equation}
As previously mentioned, here index $\alpha=A,B$ enumerates different
centers (target and projectile ions) and $\psi_{\alpha n}(\vec{r},t)$
are the wave functions of the free-moving ion $\alpha$.
After the collision ($t \to \infty$) the wave functions
$\psi_{\alpha n}(\vec{r},t)$ of the free-moving ion $\alpha$ are given by
\begin{equation}
\psi_{\alpha n}(\vec{r},t) \,=\, e^{-i E_{\alpha n}t} \,
s_{\alpha}(\vec{r}) \, \psi^{0}_{\alpha n}(\vec{r}-\vec{R}_{\alpha}) \,,
\end{equation}
where $\psi^{0}_{\alpha n}(\vec{r})$ are the stationary Dirac wave functions
of ion $\alpha$ at the rest and $s_{\alpha}(\vec{r})$ is the translation factor.
For the low-energy collisions the translation factor and the energy
$E_{\alpha n}$ of the moving ion can be taken in the nonrelativistic
approximation
\begin{equation}
s_{\alpha}(\vec{r})=\exp(i \vec{v}_{\alpha}\cdot\vec{r}) \,, \qquad
E_{\alpha n} = \varepsilon_{\alpha n}+v_{\alpha}^2/2 \,.
\end{equation}
Generally, the translation factor is introduced in the basis functions
to improve the convergence of the time-dependent wave function expansion.
We did not include the translation factor $s_{\alpha}(\vec{r})$ in the
basis functions because of the computational complexity of the
two-center integral calculations. However, at the limit $t \to \infty$
we can reexpand the moving orbitals $s_{\alpha} \varphi_{\alpha n}$ over the
basis $\varphi_{\alpha n}$ in the following way.
At the limit $t \to \infty$ the basis functions of the different centers
do not overlap. The basis set $\varphi_j$ is orthonormal and
\begin{equation}
e^{-i\vecsm{v}_{\alpha}\cdot \vecsm{r}} \, \varphi_{\alpha n}(\vec{r})
\,\simeq\,
\sum_{n^{\prime}} \, K_{\alpha n^{\prime},\alpha n} \,
\varphi_{\alpha n^{\prime}}(\vec{r}) \,, \qquad
K_{\alpha^{\prime} n^{\prime},\alpha n} = \delta_{\alpha, \alpha^{\prime}} \,
\langle \varphi_{\alpha n^{\prime}} \mid
e^{-i\vecsm{v}_{\alpha}\cdot \vecsm{r}} \mid \varphi_{\alpha n} \rangle \,.
\label{reexp1}
\end{equation}
The expansion (\ref{reexp1}) is not exact, since the finite basis
$\varphi_{\alpha n}$ is incomplete and $K$ is non-unitary matrix.
In particular,
\begin{equation}
\sum_{n^{\prime}} |\, K_{\alpha n^{\prime},\alpha n}|^2 < 1 \,.
\end{equation}
We can renormalize matrix $K$ and rewrite expansion (\ref{reexp1})
in the form
\begin{equation}
e^{-i\vecsm{v}_{\alpha}\cdot \vecsm{r}} \, \varphi_{\alpha n}(\vec{r})
\,\simeq\,
\sum_{n^{\prime}} \, \overline {K}_{\alpha n^{\prime},\alpha n} \,
\varphi_{\alpha n^{\prime}}(\vec{r}) \,, \qquad
\overline K = N \, K \,,  \qquad N= (K \, K^{+})^{-1/2} \,.
\label{reexp2}
\end{equation}
Here matrix $N$ plays a role of the normalization factor.
The renormalized matrix $\overline K$ is unitary.

Now we can obtain the expansion of the time-dependent wave
function $\Psi(\vec{r},t)$ over the basis functions with translation factor
($t \to \infty$),
\begin{equation}
\Psi(\vec{r},t) = \sum_{\alpha, n} \, C_{\alpha n}(t) \,
\varphi_{\alpha, n}(\vec{r}) \simeq \sum_{\alpha, n} \,
\overline{C}_{\alpha n}(t) \, e^{i\vecsm{v}_{\alpha}\cdot \vecsm{r}} \,
\varphi_{\alpha n}(\vec{r}) \,,
\end{equation}
where
\begin{equation}
\overline{C}_{\alpha n}(t) = 
\sum_{n^{\prime}} \, \overline{K}_{\alpha n, \alpha n^{\prime}} \,
C_{\alpha n^{\prime}}(t) \,.
\end{equation}
It should be noted that the set of coefficients $\overline C_{\alpha n}(t)$
is normalized to unity
\begin{equation}
\sum_{\alpha, n} |\overline{C}_{\alpha n}(t)|^2 = 
\sum_{\alpha, n} |{C}_{\alpha n}(t)|^2 = 1 \,, \qquad t \to \infty \,.
\end{equation}
Therefore, for the transition amplitude $T_{\alpha n}(t)$ we obtain
\begin{equation}
T_{\alpha n}(t) = \langle \psi_{\alpha n}(\vec{r},t) \mid
\Psi(\vec{r},t) \rangle =  \sum_{n^{\prime}} \,
\overline{C}_{\alpha n^{\prime}}(t) \, e^{i E_{\alpha n} \, t} \,
\langle \psi^{0}_{\alpha n} \mid \varphi_{\alpha n^{\prime}} \rangle \,,
\qquad t \to \infty \,.
\label{amplit2}
\end{equation}
The stationary Dirac wave functions $\psi^{0}_{\alpha n}$, including
wave functions of the positive and negative energy spectra, form a complete
basis set. Therefore,
\begin{equation}
\sum_{\alpha, n} |T_{\alpha n}|^2 = \sum_{\alpha, n} \,
|\overline {C}_{\alpha n}(t)|^2 =1 \,. 
\end{equation}
%
\subsubsection{Transition probabilities}
%
Transition probabilities $W_{\alpha n}(t)$ are defined by
\begin{equation}
W_{\alpha n}(t) \,=\,|T_{\alpha n}(t)|^2 \,.
\label{prob1}
\end{equation}
The probability $W_{\alpha n}(t)$, defined by equations (\ref{prob1})
and (\ref{amplit2}), has an oscillatory behavior at
$t \to \infty $, because the  basis functions
$\varphi_{\alpha,n}$ are not the solutions of the hydrogen-like Dirac
equation and the basis set is truncated. We can remove the oscillatory
component of the probability $W_{\alpha n}(t)$ for the large time
($t \to \infty$) in the same way, as it was done in Refs.
\cite{Gallaher_68, Shakeshaft_76}.

At the large time ($t \to \infty$) the coefficients $C^L_j(t)$ coincide with
the coefficients $C_j(t)$. Therefore, the coefficients $C_j(t)$ are
the solutions of equation (\ref{dirac_time})
\begin{equation}
i \, \frac{d\vec{C}(t)}{dt} \,=\, M(t) \, \vec{C}(t) \,.
\end{equation}
Then for the coefficients $\overline C_{j}(t)$ we obtain the equation
\begin{equation}
i \, \frac{d\vec{\overline C}(t)}{dt} \,=\, \overline K \, M(t) \,
\overline K^{+} \, \vec{\overline C}(t) \,.
\end{equation}
Using the diagonalization procedure for the Hermitian matrix
$\overline K \, M \, \overline K^{+}$, we can decompose
\begin{equation}
\overline K \, M \, \overline K^{+} = V \, \Omega \, V^{+},
\end{equation}
where $\Omega$ is a diagonal matrix with eigenvalues $\Omega_{kk}=\omega_k$ and
$V$ is a unitary matrix.

We introduce new coefficients $\vec{B}(t)$ by
\begin{equation}
B_k(t) \,=\, 
\left( e^{i \Omega \,t} \, V^{+} \, \vec{\overline C}(t) \right )_k =
e^{i \omega_k \,t} \, \sum_j
V^{\ast}_{jk} \, \overline C_j(t) \,, \qquad \sum_{k} |B_k(t)|^2 = 1 \,.
\end{equation}
These coefficients have  well defined limits at $t \to \infty$.
The amplitudes $T_{\alpha n}(t)$, defined by equation (\ref{amplit2}),
in terms of the coefficients $\vec{B}(t)$ are given by
\begin{equation}
T_{j}(t) = \sum_{k,l} \, e^{i (E_{j}-\omega_k) \, t} \,V_{lk} \, B_k \,
\langle \psi^{0}_{j} \mid \varphi_{l} \rangle \,.
\label{amplit3}
\end{equation}
Then for the probabilities $W_{j}(t)$ we obtain
\begin{equation}
W_{j}(t) = |T_{j}(t)|^2 = \sum_k \, \Bigl | B_k(t) \Bigr|^2 \,
\Bigl | \sum_l V_{lk} \langle \psi^{0}_j \mid \varphi_l \rangle \Bigr |^2 
\; + \; (\mbox{oscillating term}) \,.
\end{equation}
Removing the oscillating term \cite{Shakeshaft_76}, we can introduce
probabilities $W^{\prime}_{j}(t)$ defined as
\begin{equation}
W^{\prime}_{j}(t) = \sum_k \, \Bigl | B_k(t) \Bigr|^2 \,
\Bigl | \sum_l V_{lk} \langle \psi^{0}_j \mid \varphi_l \rangle \Bigr |^2 \,.
\end{equation}
Since hydrogen-like Dirac wave functions  $\psi^{0}_{\alpha,n}$
of each center $\alpha$ (including the positive and negative Dirac continuum
spectra) form a complete basis set, we get
\begin{equation}
\sum_{j \in \alpha} | W^{\prime}_{j}(t)|^2 = \sum_{k} |B(k)|^2 \,
\sum_{l} |V_{kl}|^2 = \sum_{k} |B(k)|^2 = 1 \,.
\end{equation}
The coefficients $B_k(t)$ and the matrix elements have well-defined limits
for $t \to \infty $. Therefore, there exists the limit
\begin{equation}
P_{\alpha, n} = \lim_{t \to \infty} \, W^{\prime}_{\alpha n}(t) \,.
\end{equation}
The direct ($P_{\rm d}$), charge transfer ($P_{\rm ct}$),
and ionization ($P_{\rm ion}$) probabilities are given by
\begin{equation}
P_{\rm d}   = \sumpr_{n} \, P_{A, n} \,, \qquad
P_{\rm ct}  = \sumpr_{n} \, P_{B, n} \,, \qquad
P_{\rm ion} = 1 - \sumpr_{\alpha,n} \,
P_{\alpha n} = 1 -P_{\rm d} - P_{\rm ct} \,.
\label{prob2}
\end{equation}
where the prime at the sum symbol indicates that the summation
runs over the discrete bound states of the ion $\alpha$.

The cross sections for the charge transfer ($\sigma_{\rm ct}$) and
ionization  ($\sigma_{\rm ion}$) processes are then calculated as usual
by integrating the probabilities over the impact
parameter $b$:
\begin{equation}
\sigma_{\rm ct} = 2\pi \int \limits_0^{\infty} \, db \,\, b \,
P_{\rm ct}(b) \,, \qquad
\sigma_{\rm ion} = 2\pi \int \limits_0^{\infty} \, db \,\, b \,
P_{\rm ion}(b) \,.
\label{sigma1}
\end{equation}
%
\subsubsection{Z-scaling}
It is well-known that in the nonrelativistic theory
the scale transformation $\vec{r}^{\prime}=Z \, \vec{r}$ and
$R^{\prime}=Z \, R$ allows one to transform
the wave functions $\psi(\vec{r})$ and
the energies $\varepsilon$ of a homonuclear one-electron quasi-molecule
with a point nuclear charge $Z>1$ to the wave functions
$\psi^{\prime}(\vec{r})= Z^{-3/2}\psi(Z\vec{r})$ and energies
$\varepsilon^{\prime}= \varepsilon/Z^2$ of the H$_2^{+}$ molecule.
The same scale transformation can be considered in the nonrelativistic
homonuclear collisions.

The time-dependent Schr\"odinger equation describing A$^{(Z-1)+}$-A$^{Z+}$
collision is given by
\begin{equation}
\left (- \frac{1}{2} \, \Delta +\frac{Z}{|\vec{r}-\vec{R}_A(t)|} +
\frac{Z}{|\vec{r}-\vec{R}_B(t)|} \right ) \, \Psi(\vec{r},t) \,=\,
i \frac{\partial}{\partial t} \Psi(\vec{r},t) \,.
\label{Schrod1}
\end{equation}
If in Eq. (\ref{Schrod1}) we set $\vec{r}^{\prime}=Z \, \vec{r}$,
$t^{\prime}=Z^2 \, t$, and
$\vec{R}^{\prime}_{\alpha}(t^{\prime})=Z \,\vec{R}_{\alpha}(t)$,
we obtain the time-dependent Schr\"odinger equation for the H$^{+}$-H
collision \cite{Briggs_73, Bransden_92}
\begin{equation}
\left (- \frac{1}{2} \, \Delta^{\prime} +
\frac{1}{|\vec{r}^{\prime}-\vec{R}^{\prime}_A(t^{\prime})|} +
\frac{1}{|\vec{r}^{\prime}-\vec{R}^{\prime}_B(t^{\prime})|} \right ) \,
\Psi(\vec{r}^{\prime},t^{\prime}) \,=\,
i \frac{\partial}{\partial t{^\prime}} \Psi(\vec{r}^{\prime},t^{\prime}) \,.
\label{Schrod2}
\end{equation}
It should be noted that the scaling $R^{\prime}=Z R(t)$ is satisfied
exactly for a straight-line trajectory. In this case the impact
parameter $b$, the velocity $v_{\alpha}$, and projectile energy $E$ are
transformed by
\begin{equation}
b^{\prime} = Z \, b \,, \qquad v^{\prime}_{\alpha} = v_{\alpha}/Z \,,
\qquad E^{\prime} = E/Z^2\,.
\end{equation}
It follows that the probability $P_{\alpha,n}(Z,b,E)$ 
and the cross section $\sigma_{\alpha,n}(Z,E)$
for a process in a
symmetric system with nuclei of charge Z can be obtained from the probability
$P_{\alpha,n}(1,b^{\prime},E^{\prime})$
and the cross section $\sigma_{\alpha,n}(1,E^{\prime})$
for the same process in the H($1s$)--H$^{+}$ system by the relations
\begin{equation}
P_{\alpha,n}(Z,b,E)=P_{\alpha,n}(1,b\,Z ,E/Z^2) \,,
\qquad
\sigma_{\alpha,n}(Z,E)= \frac{1}{Z^2} \, \sigma_{\alpha,n}(1,E/Z^2) \,.
\label{scale1}
\end{equation}
It should be noted that this scaling law is not valid for 
the relativistic collisions.
%
\section{Results}
\label{sec:results}
\subsection{The choice of the basis}
\label{basis1:}
In our relativistic calculations, we used two Dirac-Sturm bases of
different size. Both bases include functions of the positive-energy
Dirac spectrum and Sturm orbitals corresponding to the negative-energy Dirac
spectrum.  It should be noted that the constructed bases satisfy
the dual kinetic balance conditions \cite{Shab_Tup_04} and do not contain
so-called ``spurious'' states \cite{Tup_Shab_08}.

The positive-energy functions of the first basis (Basis 1) on
each center
in the standard nonrelativistic notations of atomic shells are given by:
1s-3s, 2p,3p, 3d,
$\overline{4\rm s}$-$\overline{6\rm s}$,
$\overline{4\rm p}$-$\overline{6\rm p}$,
$\overline{4\rm d}$-$\overline{6\rm d}$,
$\overline{4\rm f}$,$\overline{5\rm f}$.
Here the overline symbol ($\overline{\rm nl}$) is used to indicate
the Dirac-Sturm
(pseudo-state) basis functions. The total number of the positive-energy
orbitals of both centers is $220$ and the total size of Basis 1, including
the negative-energy orbitals, is $440$. Basis 1 is used
for both stationary and time-dependent wave functions.

The basis size can be increased
in the calculations of the stationary states of quasi-molecules. Positive-energy
functions of the second basis (Basis 2) is constructed from $26$
atomic shells:
1s, $~\overline{2\rm s}$-$\overline{8\rm s}$,
$~\overline{2\rm p}$-$~\overline{8\rm p}$,
$~\overline{3\rm d}$-$~\overline{8\rm d}$
$~\overline{4\rm f}$-$~\overline{6\rm f}$
$~\overline{5\rm g}$,$~\overline{6\rm g}$.
In Basis 2, the total number of orbitals of both ions, including the
negative-energy
spectrum, is equal to $784$. This basis is used only in
the calculations of the stationary states of quasi-molecules.
\subsection{Stationary ground states of some homonuclear quasi-molecules}
\label{subsec:stationary}
\subsubsection{Energies of the ground state of some homonuclear quasi-molecules}
In  Table \ref{Mol_energies} we present the results of  our
relativistic calculations of the $1\sigma_{g}$ state energy of the H$_2^{+}$,
Th$_2^{179+}$, and U$_2^{183+}$ quasi-molecules for so-called chemical distance
$R=2/Z$ a.u.. Since the quasi-molecule Th$_2^{179+}$
was considered as a reference system
for testing relativistic
effects, it was calculated in a number of papers  using high-precision
large-scale methods (see, e.g., Refs.
\cite{Parpia_95, LaJohn_98, Rutkowski_99, Kullie_01}).
\begin{table}[ht]
\caption{Relativistic energies (a.u.) of the 1$\sigma_{g}$ state of
quasi-molecules for the point-charge nuclei and  $R=2/Z$ a.u..}
\vspace{1.0mm}
\begin{center}
\vspace{-2.0mm}
\begin{tabular}{|c|c|c|c|c|c|}
\hline
\multicolumn{1}{|c|}{} & \multicolumn{2}{|c|}{} & \multicolumn{2}{c|}{}
 & \multicolumn{1}{c|}{} \\[-6mm]
\multicolumn{1}{|c|}{} & \multicolumn{2}{|c|}{H$_2^+$ $(Z=1)$}
& \multicolumn{2}{c|}{Th$_2^{179+}$ $(Z=90)$}
& \multicolumn{1}{c|}{U$_2^{183+}$ $(Z=92)$} \\[0.0mm]
\hline &&&&&\\[-5.5mm]
& $\vec{\varepsilon_{1\sigma_{+}}}$ & Rel. error
& $\vec{\varepsilon_{1\sigma_{+}}}$ & Rel. error
& $\vec{\varepsilon_{1\sigma_{+}}}$   \\[0mm]
\hline &&&&&\\[-4.5mm]
~Basis 1~ & $-1.1026248$~    & ~$1.5 \cdot 10^{-5}$~ & $-9504.573$~~
                           & ~$1.9 \cdot 10^{-5}$~ & $-9965.190$~~ \\[-1mm]
~Basis 2~ & $-1.1026405$~    & ~$1.0 \cdot 10^{-6}$~ & $-9504.732$~~
                           & ~$2.5 \cdot 10^{-6}$~ & $-9965.307$~~ \\[-1mm]
 Others    & $-1.1026416^a$ &                       & $-9504.756^b$ &   &
\\[1mm]
\hline
\multicolumn{6}{l}{}\\[-6.5mm]
\multicolumn{6}{l}{$^a$ Ref. \cite{Yang_91}}\\[-2mm]
\multicolumn{6}{l}{$^b$ Ref. \cite{Kullie_01}} \\[0mm]
\end{tabular}
\end{center}
\label{Mol_energies}
\end{table}
As one can seen from Table \ref{Mol_energies}, there is a good agreement of
our data with very accurate  values obtained in Refs. \cite{Yang_91,Kullie_01}.
The relative precision of our results for the quasi-molecules H$_2^{+}$
and Th$_2^{179+}$  is increased
by an order of magnitude when Basis 1 is replaced by Basis 2.

\begin{figure}[ht]
\centering
\includegraphics[width=11cm,clip]{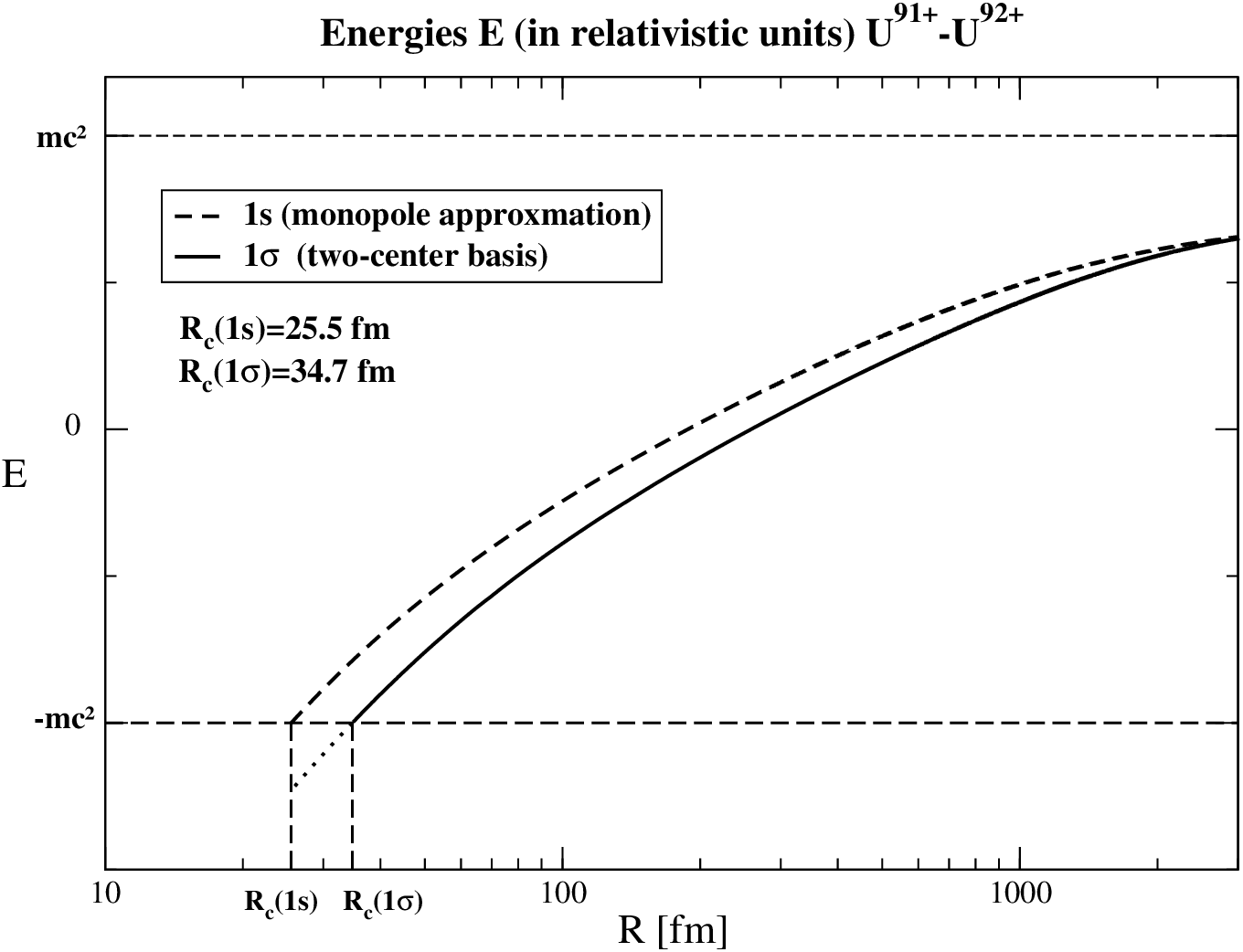}
\vspace{-6mm}
\caption{\small The $1\sigma_g$ state energy of the U$_2$ quasi-molecule as a
function of the internuclear distance $R$ on a logarithmic scale.}
\label{U2_energies}
\end{figure}

In Fig. \ref{U2_energies} we display the energy of 
the $1\sigma_g$ state of the U$_2^{183+}$ quasi-molecule as a function
of the internuclear distance $R$ 
on a logarithmic scale.
In this figure the solid line
indicates the energy
$E(R)$ calculated using two-center Dirac-Sturm Basis 2. The dashed
line represent the results of the one-center calculations in the monopole
approximation. As one can see from Fig. \ref{U2_energies},
in the two-center basis
the 1$\sigma_g$ electron "dives" into the
negative-energy Dirac continuum at a critical distance 
$R_{\rm c} = 34.7$ fm.  The critical distance obtained in our monopole
approximation amounts to $R_{\rm c}=25.5$ fm that is too far from the
exact value. 
It should be noted that in our one-center monopole approximation
the basis was centered at the position of the nucleus ($A$ or $B$) but not at
the center of the internuclear interval, as it was done in
the papers \cite{Muller_72, Muller_76, Soff_78}. The monopole
approximation used in Ref. \cite{Soff_78} is more suitable
for the short-distance regime and gives the value of the critical
distance equal to $R_{\rm c} = 35$ fm.
%
\subsubsection{Critical distance}
\label{subsec:critical}
In the Table \ref{crit_rad} we present our results of the two-center
relativistic calculations (Basis 2) of the critical distance $R_{\rm c}$ 
for 
homonuclear one-electron quasi-molecules A$_2^{(2Z-1)+}$ and compare
them with the corresponding values obtained by other authors. There exists a 
discrepancy of about $5$-$10\%$ between the critical distance data for
the point nuclei
\cite{Marinov_75, Rafelski_76, Lisin_77, Soff_78, Wietschorke_79, Matveev_00}.
Our results for this case are in a very good agreement with the
results of Ref. \cite{Lisin_77}.
\begin{table}[ht]
\caption{Critical distances $R_c$ (fm) for homonuclear one-electron
quasi-molecules A$_2^{(2Z-1)+}$.}
\vspace{-4.0mm}
\begin{center}
\begin{tabular}{|c|c|c|c|c|c|}
\hline
\multicolumn{1}{|c|}{} & \multicolumn{2}{|c|}{} & \multicolumn{3}{|c|}{}\\[-5mm]
\multicolumn{1}{|c|}{} & \multicolumn{2}{|c|}{Point nucleus} &
\multicolumn{3}{|c|}{Extended nucleus}\\[2mm]
\hline &&&&& \\[-5mm]
~~~$Z$~~~ & ~This work~ & ~~~Others~~~ & $\langle R^2_{\rm n} \rangle^{1/2}$(fm)
  & ~This work~ & ~Others~ \\[2mm]
\hline &&&&&\\[-5.5mm]
 88  & 24.27 &  24.24$^a$    & 5.5705 &  19.91  &   19.4$^{c}$  \\[-1mm]
 90  & 30.96 &  30.96$^a$    & 5.7210 &  27.05  &   26.5$^{c}$  \\[-1mm]
 92  & 38.43 &  38.42$^a$    & 5.8569 &  34.72  &   34.3$^{c}$  \\[-2mm]
     &       &  36.8$^{~b}$~ &        &         &   34.7$^{d}$  \\[-1mm]
 94  & 46.58 &  46.57$^a$    & 5.794  &  43.16  &   42.6$^{c}$  \\[-1mm]
 96  & 55.38 &  55.37$^a$    & 5.816  &  52.09  &   51.6$^{c}$  \\[-1mm]
 98  & 64.79 &  64.79$^a$    & 5.844  &  61.63  &   61.0$^{c}$  \\[-2mm]
     &       &               &        &         &   61.1$^{d}$  \\[-1mm]
\hline
\multicolumn{6}{l}{}\\[-6mm]
\multicolumn{6}{l}{$^a$ Ref. \cite{Lisin_77}, $^b$ Ref. \cite{Rafelski_76},
$^c$ Ref. \cite{Lisin_80}, $^d$ Ref. \cite{Muller_76}} \\[-6mm]
\end{tabular}
%
%
\end{center}
\label{crit_rad}
\end{table}
In our calculations for extended nuclei, the finite nuclear size was
taken into account
using the Fermi model of the nuclear charge distribution
(for details, see, e.g., Ref. \cite{Parpia_92}).
The root-mean-square nuclear charge radii
$\langle R^2_{\rm n} \rangle^{1/2}$ were taken from Refs. \cite{Angeli_04}
(for $Z$=$88$), \cite{Kozhedub-Shabaev_08} (for $Z$=$90$),
\cite{Kozhedub_08} (for $Z$=$92$), and \cite{Johnson-Soff_85} (for $Z$=$94$,
$96$, $98$).
The number of works where the finite nuclear size effect was taken into
account is much less than for the point nucleus case. We can systematically
compare our results only with the data obtained in Ref. \cite{Lisin_80}.
The discrepancy between our data and those from Ref. \cite{Lisin_80}
is considerably larger for the extended nuclei than for the point nuclei.
A possible reason of that could consist in a rather crude estimate of
the nuclear size effect in Ref. \cite{Lisin_80}. 
It should also be noted that in the work \cite{Lisin_80} other
values of the nuclear radii, namely
$\langle R^2_{\rm n} \rangle^{1/2}=\sqrt{3/5} \cdot 1.2 \cdot
(2.6 \cdot Z)^{1/3}$ were used. Our calculations showed, however, that the usage
of the nuclear radii from Ref. \cite{Lisin_80} changes the values of
$R_{\rm cr}$ not more than by $0.02$ fm.
\subsection{Charge-transfer probabilities and cross sections}
\label{subsec:Charge-trnasfer}
\subsubsection{H(1s)--H$^{+}$ collisions}
\label{subsec:H2}
\begin{figure}
\centering
\includegraphics[width=12cm, clip]{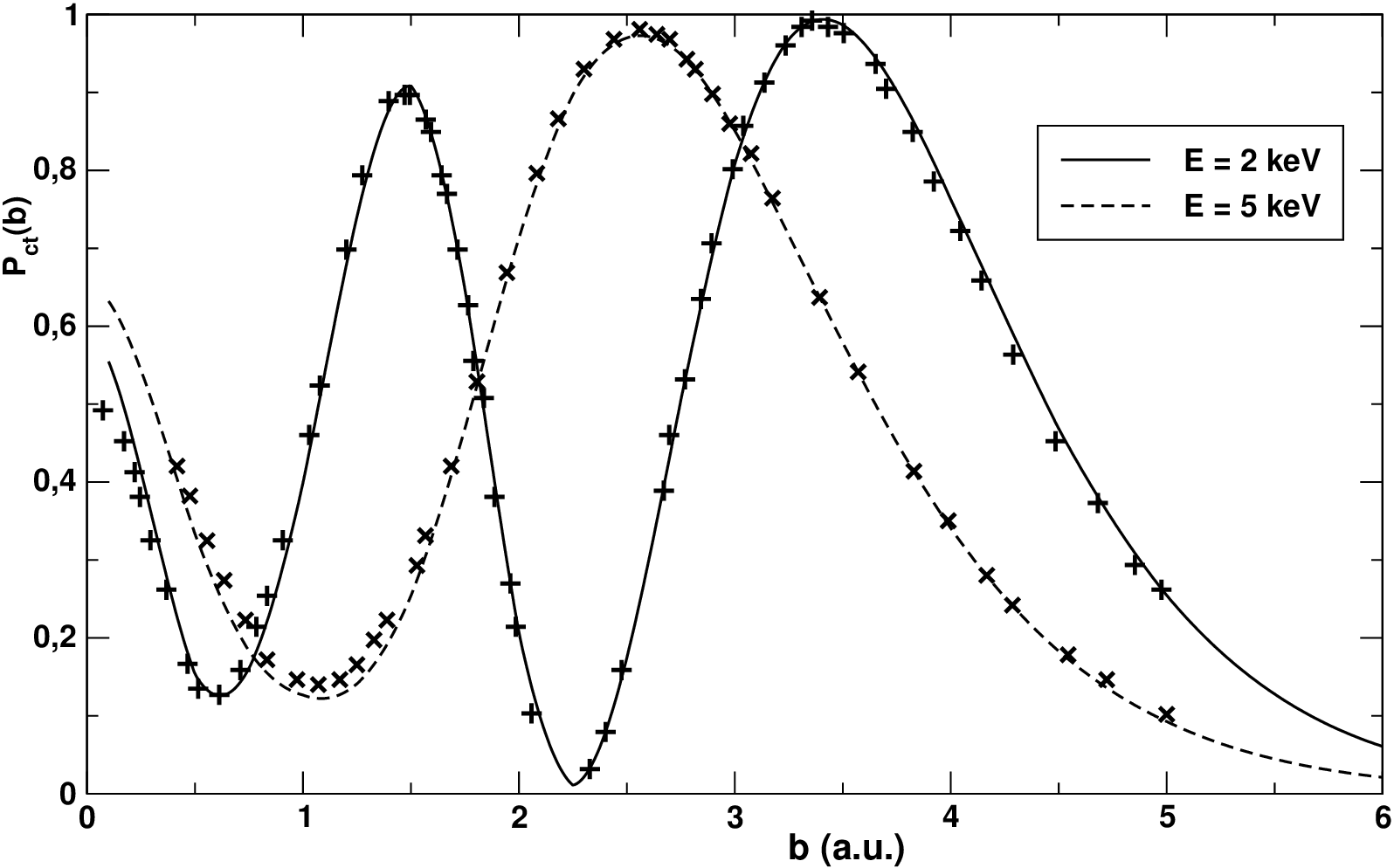}
\caption{\small Charge-transfer probabilities P$_{\rm ch}$(b) for
the H(1s)-H$^{+}$ collision as functions of the impact parameter $b$.
Our results for the projectile energies $2$ keV  (solid line) and
$5$ keV (dashed line) are compared to the related results from the AO-expansion
calculations \cite{Fritsch_82, Fritsch_91} (symbols "$+$" and "$\times$").}
\label{H2_prob}
\end{figure}

Fig. \ref{H2_prob} shows the
charge-transfer probabilities $P_{ct}(b)$ for the H$(1s)$--H$^{+}$ collision
as functions of the impact parameter $b$ for
the projectile energies of $2$ keV and $5$ keV.
The results of our relativistic calculations for $2$ keV (solid line)
and $5$ keV (dashed line) are found to be very close to nonrelativistic
calculations based on the 
 two-center atomic-orbital (AO) expansion \cite{ Fritsch_82, Fritsch_91}.
This is not surprising, since the relativistic effects are negligible for the
H$(1s)$--H$^{+}$
collision. Our calculations were performed for the straight-line
trajectory
of the projectile that corresponds to 
the full screening of the
target nuclear charge by the $1s$-electron.

 As was demonstrated in Ref. \cite{Fritsch_91}, the two-center AO
expansion data are in a very good agreement with results 
obtained by a direct  numerical solution of the
nonrelativistic Schr\"odinger equation
\cite{Grun_82}. In the review \cite {Fritsch_91},
 the two-center AO expansion data \cite{Fritsch_82} are also compared
with the results \cite{Ludde_81, Ludde_82}, obtained by the expansion
over the ``nonmoving'' Hylleraas functions. It should be noted
that the Hylleraas expansion data are similar to the results of
Refs. \cite{ Fritsch_82},
\cite{Grun_82} and to our results in trend but differ
by the phase and the magnitude.

In Table \ref{H2_cross} we present the total charge-transfer
cross sections $\sigma_{\rm ct}$ for the H($1s$)-H$^{+}$ collision
in a wide range of the projectile energy (from $0.5$ keV to $100$ keV)
and  compare them  with nonrelativistic large-scale calculations of the
recent paper \cite{Winter_09}, which can be considered as
an extension of the pioneering works \cite{Shakeshaft_76, Shakeshaft_78},
where the analytical Sturmian basis set expansion was used.
We also give the cross section values, 
deduced from the experimental results \cite{Janev_88}. The relative
uncertainties  of the recommended and interpolated experimental 
data are about $5\%$-$10\%$.
As one can see from the table \ref{H2_cross}, our results
are in a good agreement with the theoretical data of Ref. \cite{Winter_09}
and with the experimental data.
%
\begin{table}[ht!]
\small
\caption{Charge transfer cross section  $\sigma_{\rm ct}$ for the
H$(1s)$-H$^{+}$ collision, in units of $10^{-17}$ cm$^2$.}
\vspace{-3.0mm}
\begin{center}
%
\begin{tabular}{|c|c|c|c|}
\hline &&& \\[-4mm]
Projectile energy &  ~~~~~~$\sigma_{\rm ct}(E)$~~~~~~~&  ~~~~$\sigma_{\rm ct}(E)$~~~~ &
$\sigma_{\rm ct}(E)$  \\[1mm]
~~~E (keV)~~~ & ~~~~~This work~~~~~ & ~~Winter \cite{Winter_09}~~ &
~~~~~Exp$^{a}$~~~~  \\[0.5mm]
\hline &&&\\[-3.5mm]
  0.5    & 199.6 &          &                       \\[-1.5mm]
  0.7    & 186.9 &          &                       \\[-1.5mm]
  1.0  & 172.4   &  173.0   & 171~                  \\[-1.5mm]
  2.0  & 144.9   &          & 144~                  \\[-1.5mm]
  4.0  & 117.5   &  118.1   & 115$^{\ast}$          \\[-1.5mm]
  5.0  & 107.8   &          & 110~                  \\[-1.5mm]
 10.0  &  81.3   &          &  77.5~                \\[-1.5mm]
 15.0  &  63.5   &  67.41   &  55.6$^{\ast}$        \\[-1.5mm]
 20.0  &  48.9   &          &  44.5~                \\[-1.5mm]
 25.0  &  36.2   &  39.45   &  35.3$^{\ast}$        \\[-1.5mm]
 30.0  &  26.6   &          &  27.6$^{\ast}$        \\[-1.5mm]
 40.0  &  15.3   &          &  16.5$^{\ast}$        \\[-1.5mm]
 50.0  &   9.1   &  10.04   &   9.9                 \\[-1.5mm]
 60.0  &   5.6   &          &   5.9$^{\ast}$        \\[-1.5mm]
 70.0  &   3.5   &          &   3.6$^{\ast}$        \\[-1.5mm]
 80.0  &   2.3   &          &   2.3$^{\ast}$        \\[-1.5mm]
100.0  &   1.1   &   1.11   &   1.1~                \\[ 0.5mm]
\hline
\multicolumn{4}{l}{}\\[-5mm]
\multicolumn{4}{l}{$^a$ Recommended values \cite{Janev_88} deduced from
the experimental data.} \\[-1mm]
\multicolumn{4}{l}{$^{\ast}$ Interpolated values obtained using an analityc fitting
function \cite{Janev_88}.} \\[-1mm]
\end{tabular}
%
%
\end{center}
\label{H2_cross}
\end{table}

\begin{figure}
\centering
\includegraphics[width=12cm, clip]{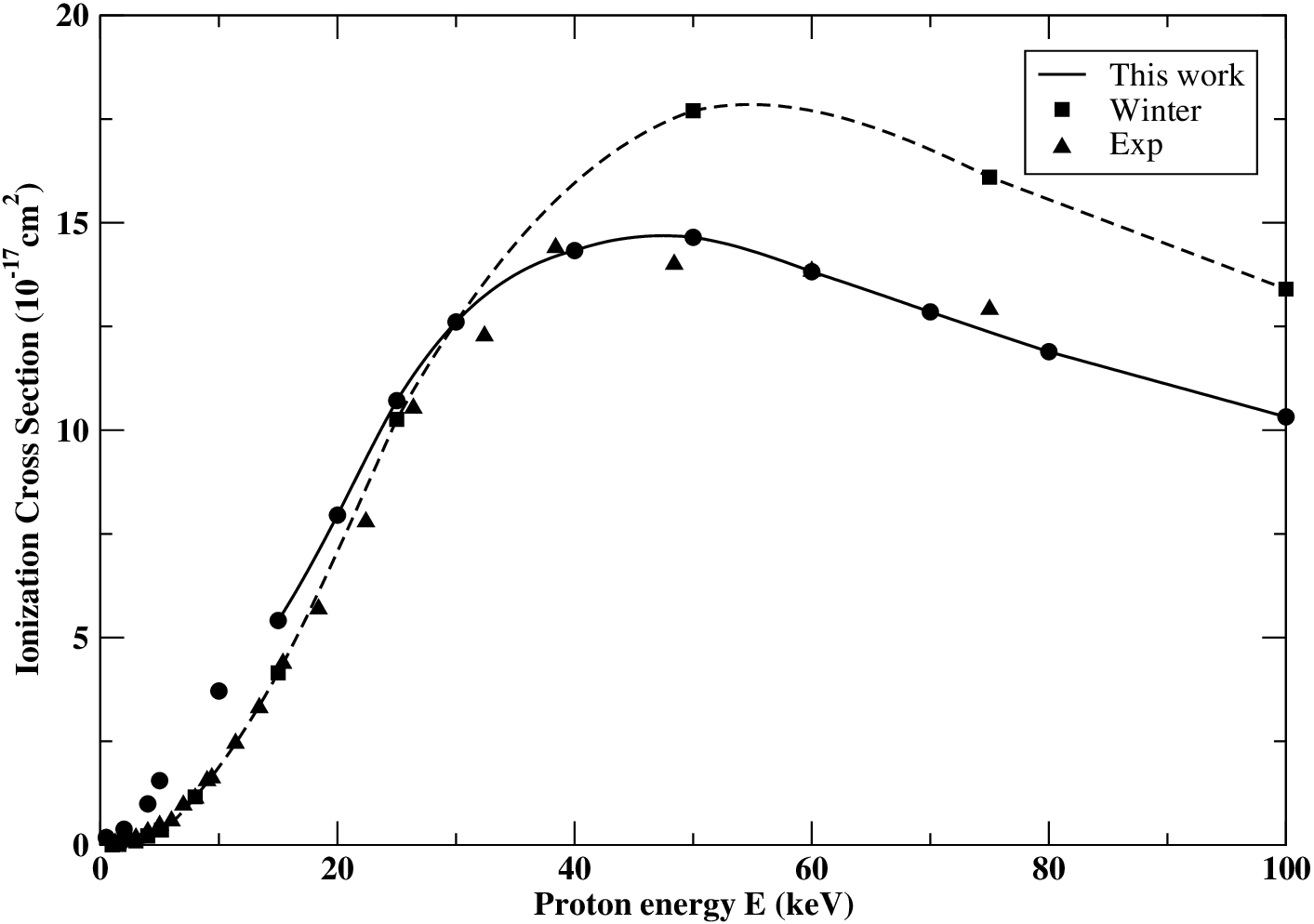}
\caption{\small Ionization cross section $\sigma_{\rm ion}$(E) for
the H$(1s)$-H$^{+}$ collision as a function of the projectile
energy $E$. The solid line is obtained by the 
interpolation of  our results indicated by 
circles, the dashed line is obtained by the
interpolation of data from Ref. \cite{Winter_09}, 
and the triangles indicate the experimental data from
Refs. \cite{Shah_87, Shah_98}.}
\label{H2_ionization}
\end{figure}

The ionization cross sections, computed in this work using
equations (\ref{prob2}) and (\ref{sigma1}), are displayed in
Fig.~\ref{H2_ionization}.
Our results are in a good agreement with the experimental data
in the range of the proton energy from $20$ keV to $80$ keV. 
At the energies
less than $15$ keV we observe a significant relative deviation of our results 
from the experimental
data. 
Thus is probably due to the fact that the absolute
uncertainty of
our data is approximately the same  in the whole region of the energies
( about (1\textendash3)$\cdot 10^{-17}$~cm$^2$), while 
at the low energies the ionization cross section tends to
zero. This leads to a large relative error in the low energy region.

In contrast to our results, the theoretical data of Ref. \cite{Winter_09}, 
 which are shown in Fig.~\ref{H2_ionization} by squares,
 are in a good agreement with the experimental
data in the low energy region (less than $25$ keV) and significantly
differ (at least by $25\%$) from the experimental data for the energies
larger than $40$ keV. The reason of this discrepancy is unclear to us.
%
\subsubsection{Ne$^{9+}$(1s)--Ne$^{10+}$ collisions}\label{subsec:Ne2}
To study the role of the relativistic effects in the homonuclear
collisions and to test our approach we calculated the charge transfer
cross sections for the Ne$^{9+}(1s)$--Ne$^{10+}$ collisions with the
standard value of the speed of light ($c=137.036$ a.u.)
and in the nonrelativistic limit ($c\to \infty$) by multiplying the
standard value of the speed of light by the factor $1000$. The obtained values
are presented in Table~\ref{Ne2_cross}.
\begin{table}[ht!]
\small
\caption{Charge transfer cross section $\sigma_{\rm ct}(E)$
($10^{-17}$ cm$^2$) as a function of the projectile energy E
for the Ne$^{9+}(1s)$-Ne$^{10+}$ ($Z$=$10$) and H$(1s)$-H$^{+}$ collisions.}
\vspace{-2.0mm}
\begin{center}
%
\begin{tabular}{|c|c|c|c|c|}
\hline
\multicolumn{1}{|c|}{} & \multicolumn{2}{|c|}{} &
\multicolumn{1}{c|}{}& \multicolumn{1}{c|}{} \\[-5mm]
\multicolumn{1}{|c|}{} &
\multicolumn{2}{|c|} {Ne$^{9+}(1s)$-Ne$^{10+}$} &
\multicolumn{1}{c|}{~H$(1s)$-H$^{+}$ ~} &
\multicolumn{1}{|c|} {Ne$^{9+}(1s)$-Ne$^{10+}$}
\\[1mm]
\hline &&&& \\[-5mm]
$E/Z^2$ &  ~~$\sigma_{\rm ct}(E) \cdot Z^2$~~ &
          ~~$\sigma_{\rm ct}(E) \cdot Z^2$~~ &
          ~~~~$\sigma_{\rm ct}(E)$~~~~  &
          ~~$\sigma_{\rm ct}(E) \cdot Z^2$~~ \\[-2mm]
~(keV/u)~ & Rel.$^a$ & Nonrel.$^b$ & & Born approximation$^c$ \\[2mm]
\hline &&&&\\[-5.5mm]
 1.0   &  171.6 &  172.2  &  172.4  & 188.4  \\[-1.5mm]
 2.0   &  144.3 &  144.8  &  144.9  & 150.7  \\[-1.5mm]
 4.0   &  117.1 &  117.5  &  117.5  & 114.8  \\[-1.5mm]
 5.0   &  107.3 &  107.7  &  107.8  & 107.3  \\[-1.5mm]
 10.0  &   80.8 &   81.3  &   81.3  &  76.2 \\[-1.5mm]
 15.0  &   63.0 &   63.5  &   63.5  &  57.6 \\[-1.5mm]
 20.0  &   48.5 &   48.9  &   48.9  &  48.2 \\[-1.5mm]
 25.0  &   35.9 &   36.2  &   36.2  &  38.1 \\[-1.5mm]
 30.0  &   26.4 &   26.7  &   26.6  &  30.1 \\[-1.5mm]
 40.0  &   15.1 &   15.3  &   15.3  &  19.9 \\[-1.5mm]
 50.0  &   9.0 &     9.1  &    9.1  &  13.7 \\[-1.5mm]
 60.0  &   5.6 &     5.6  &    5.6  &   9.1 \\[-1.5mm]
 70.0  &   3.5 &     3.5  &    3.5  &   5.4 \\[-1.5mm]
 80.0  &   2.3 &     2.3  &    2.3  &   3.6 \\[-1.5mm]
100.0  &   1.1 &     1.1  &    1.1  &   2.0 \\[0.5mm]
\hline
\multicolumn{5}{l}{}\\[-6mm]
\multicolumn{5}{l}{$^a$ Relativistic calculations.} \\[-2mm]
\multicolumn{5}{l}{$^b$ Nonrelativistic limit ($c\to \infty$).} \\[-2mm]
\multicolumn{5}{l}{$^c$ Born approximation \cite{Shevelko_09}.}
\\[0mm]
\end{tabular}
%
%
\end{center}
\label{Ne2_cross}
\end{table}
It should be noted that the projectile energy values are divided
by  $Z^2$ ($Z$ is the nuclear charge) and the values of
the ionization cross section $\sigma_{\rm ion}$ are multiplied by the
factor $Z^2$. This was done in order to compare the
Ne$^{9+}$($1s$)--Ne$^{10+}$ cross section
data with the H($1s$)--H$^{+}$ results in accordance with the scaling low
(\ref{scale1}).
As one can see from the table, the relativistic effects, which
decrease the values of the charge-transfer cross section, are
rather small and can be estimated as $0.5$-$0.8$ $(\alpha Z)^2$, where
$\alpha$ is the fine-structure constant.
In Table \ref{Ne2_cross}, we also compare our scaled nonrelativistic
Ne$^{9+}$($1s$)--Ne$^{10+}$ data with the H($1s$)--H$^{+}$ results.
It should be noted that our calculations for the Ne$^{9+}$($1s$)--Ne$^{10+}$
collision were performed for the Rutherford trajectory
(see section \ref{subsubsec:rutherford_sec}). This is probably the
reason of a very small discrepancy between the data presented in
the 3-rd and 4-th columns of Table \ref{Ne2_cross}.

It is also of interest to compare our results with the calculations performed
within the plane wave Born (PWB) approximation. The results of such a
calculation for the Ne$^{9+}$($1s$)--Ne$^{10+}$ collision \cite{Shevelko_09} are
presented in the 5-th column of Table \ref{Ne2_cross}. The details of the
modified PWB method can be found in Ref. \cite{Shevelko_01}. It is seen from
the table, that the PWB data are in a reasonable agreement with the more
elaborated calculation.

\subsubsection{Xe$^{53+}$(1s)--Xe$^{54+}$ collisions}\label{subsec:Xe2}
\begin{figure}
\centering
\includegraphics[width=12cm, clip]{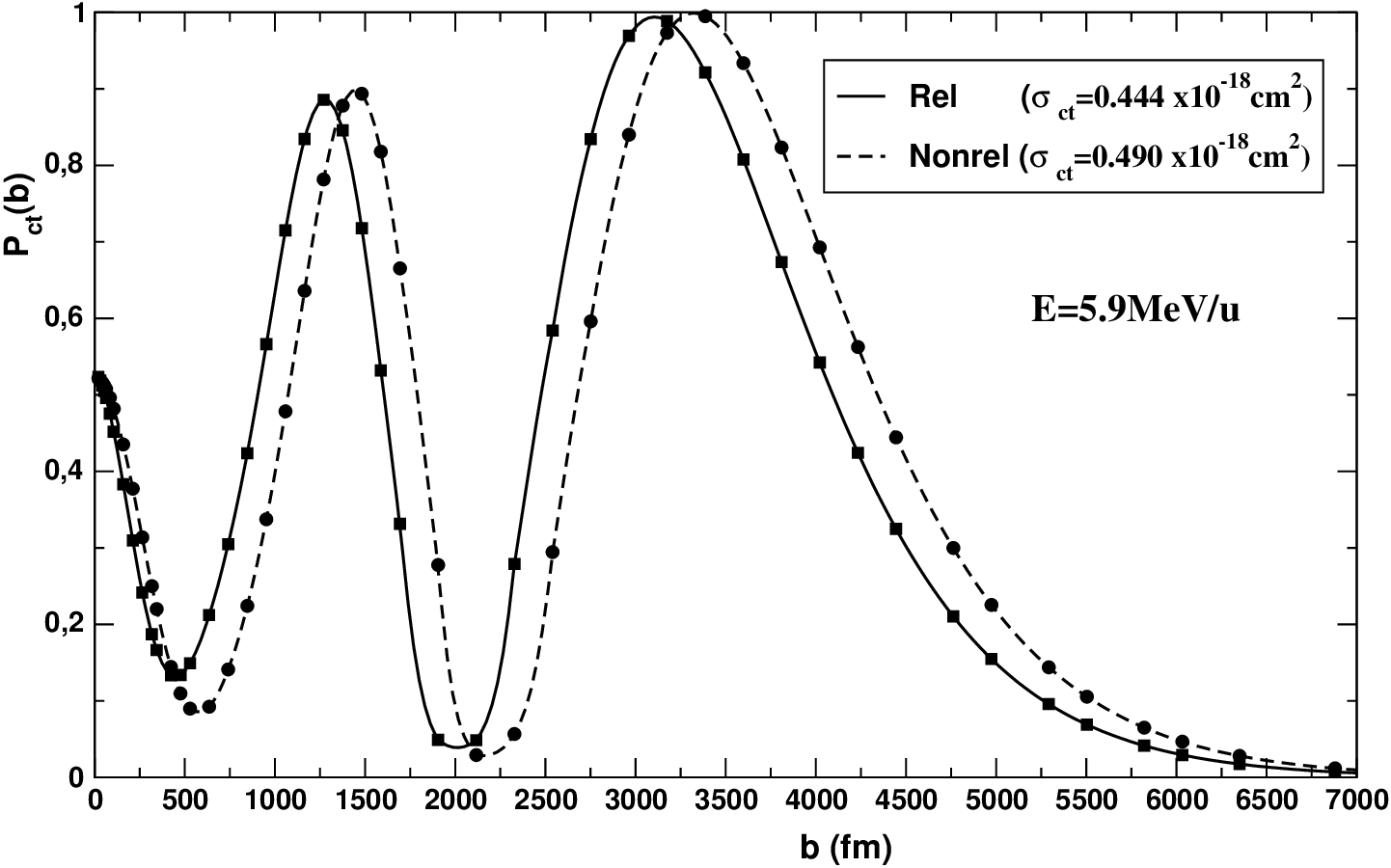}
\caption{\small The charge-transfer probability $P_{\rm ch}(b)$ for
the  Xe$^{53+}$($1s$)--Xe$^{54+}$ collision as a function of the impact
parameter $b$. The solid line interpolates the relativistic
values (squares) and the dashed line corresponds to the nonrelativistic
limit. In both cases, the projectile energy is $E$=$5.9$ MeV/u.}
\label{Xe2_prob}
\end{figure}

The relativistic effect for the Xe$^{53+}$($1s$)--Xe$^{54+}$ collisions
is considerably larger than for the Ne$^{9+}$($1s$)--Ne$^{10+}$
collisions. The computed relativistic (solid line) and nonrelativistic
(dashed line) charge transfer probabilities $P_{\rm ct}(b)$ as functions
of the impact parameter $b$ for the projectile energy of $5.9$ MeV/u are
displayed in Fig.~\ref{Xe2_prob}. The oscillatory behavior of both
 curves is the same but the nonrelativistic
curve is shifted toward higher energies. 

In Table \ref{Xe2_cross} we present the relativistic and 
non-relativistic values of the charge-transfer cross section
for the  Xe$^{53+}$($1s$)--Xe$^{54+}$ collision
scaled to $Z=1$.
One can see from the table, the relativistic effect increases from
$10\%$
to $40\%$ with increasing the projectile energy.

\begin{table}[ht!]
\small
\caption{The charge transfer cross section $\sigma_{\rm ct}(E)$
($10^{-17}$ cm$^2$) as a function of the projectile energy $E$
for the Xe$^{9+}$($1s$)-Xe$^{10+}$ and H($1s$)-H$^{+}$ collisions.}
\vspace{-2.0mm}
\begin{center}
%
\begin{tabular}{|c|c|c|c|c|}
\hline
\multicolumn{1}{|c|}{} & \multicolumn{3}{|c|}{} &
\multicolumn{1}{c|}{}\\[-5mm]
\multicolumn{1}{|c|}{} & \multicolumn{3}{|c|}
{Xe$^{53+}$($1s$)-Xe$^{54+}$ ($Z$=$54$)} &
\multicolumn{1}{c|}{~H($1s$)-H$^{+}$ ($Z$=$1$)~}\\[1mm]
\hline &&&& \\[-5mm]
$E/Z^2$ & E & ~~$\sigma_{\rm ct}(E) \cdot Z^2$~~ &
          ~~$\sigma_{\rm ct}(E) \cdot Z^2$~~ &
          ~~~~$\sigma_{\rm ct}(E)$~~~~  \\[-2mm]
 ~(keV/u)~ & ~(MeV/u)~ & Rel.$^a$ & Nonrel.$^b$ &  \\[2mm]
\hline &&&&\\[-5.5mm]
1.23457 &   3.6   &  148.3  &  163.3  &  165.0     \\[-1.5mm]
2.02332 &   5.9   &  129.4  &  143.0  &  144.9     \\[-1.5mm]
3.42936 &  10.0   &  109.1  &  123.8  &  124.8     \\[-1.5mm]
34.2936 & 100.0  &    13.3  &   20.6  &   20.7     \\[0.5mm]
\hline
\multicolumn{5}{l}{}\\[-6mm]
\multicolumn{5}{l}{$^a$ Relativistic calculations.} \\[-2mm]
\multicolumn{5}{l}{$^b$ Nonrelativistic limit ($c\to \infty$).} \\[0mm]
\end{tabular}
%
%
\end{center}
\label{Xe2_cross}
\end{table}
\subsubsection{U$^{91+}$(1s)--U$^{92+}$ collisions}
\label{subsec:U2}
%
\begin{figure}
\centering
\includegraphics[width=12cm, clip]{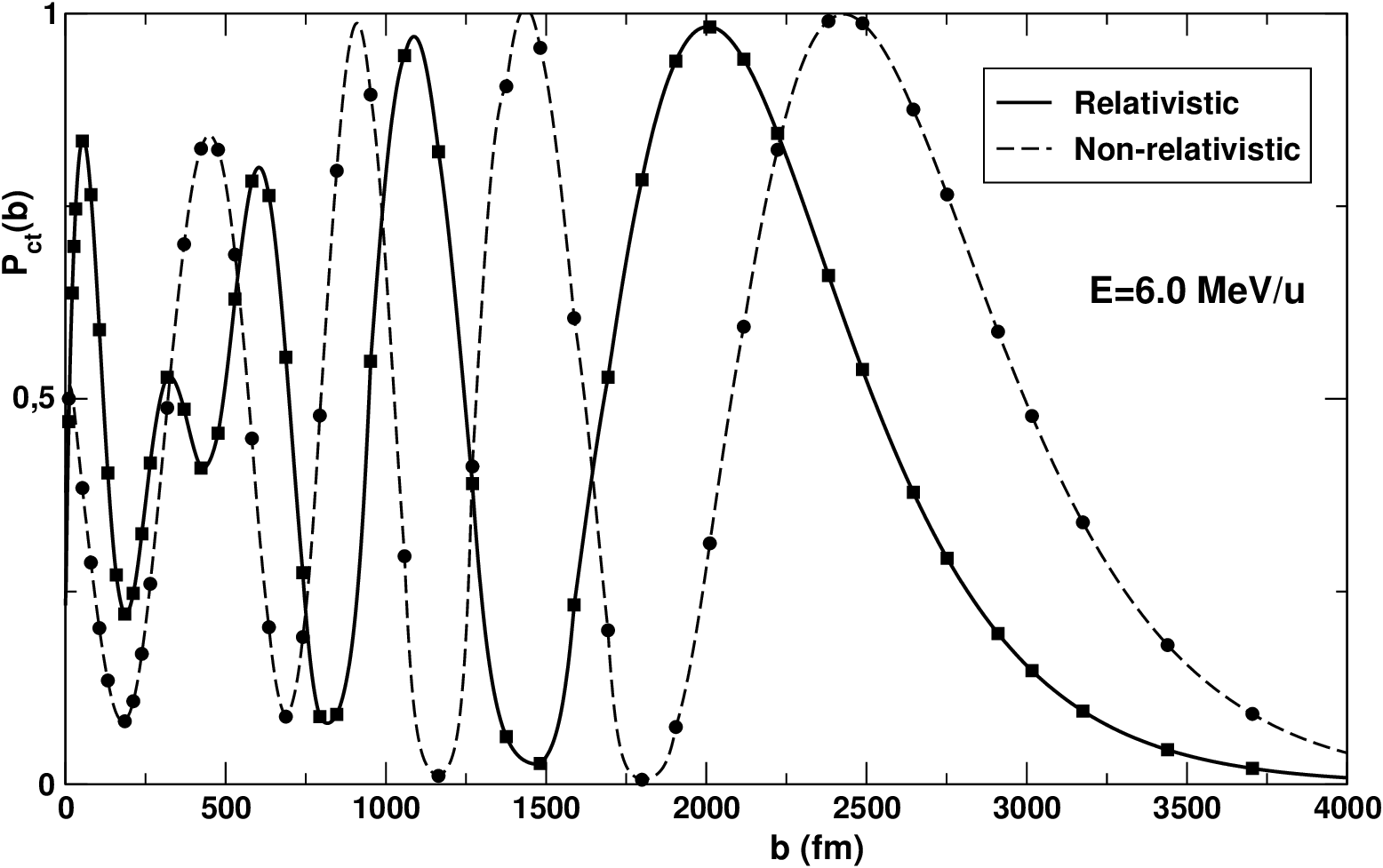}
\caption {
Charge-transfer probability $P_{\rm ch}(b)$ for
the  U$^{91+}$($1s$)--U$^{92+}$ collision as a function of the impact
parameter $b$. The solid line interpolates the relativistic
values (squares) and the dashed line corresponds to the nonrelativistic
limit. In both cases, the projectile energy is $E=6.0$ MeV/u.}
\label{U2_prob}
\end{figure}
%
The calculations of the charge-transfer probabilities and cross sections
for the U$^{91+}$($1s$)--U$^{92+}$ collisions
were performed for the extended nuclei. 
The Fermi model of the nuclear charge distribution with 
$R_{\rm nucl}=5.8569$~fm was used \cite{Kozhedub_08}.

\begin{figure}
\centering
\includegraphics[width=12cm, clip]{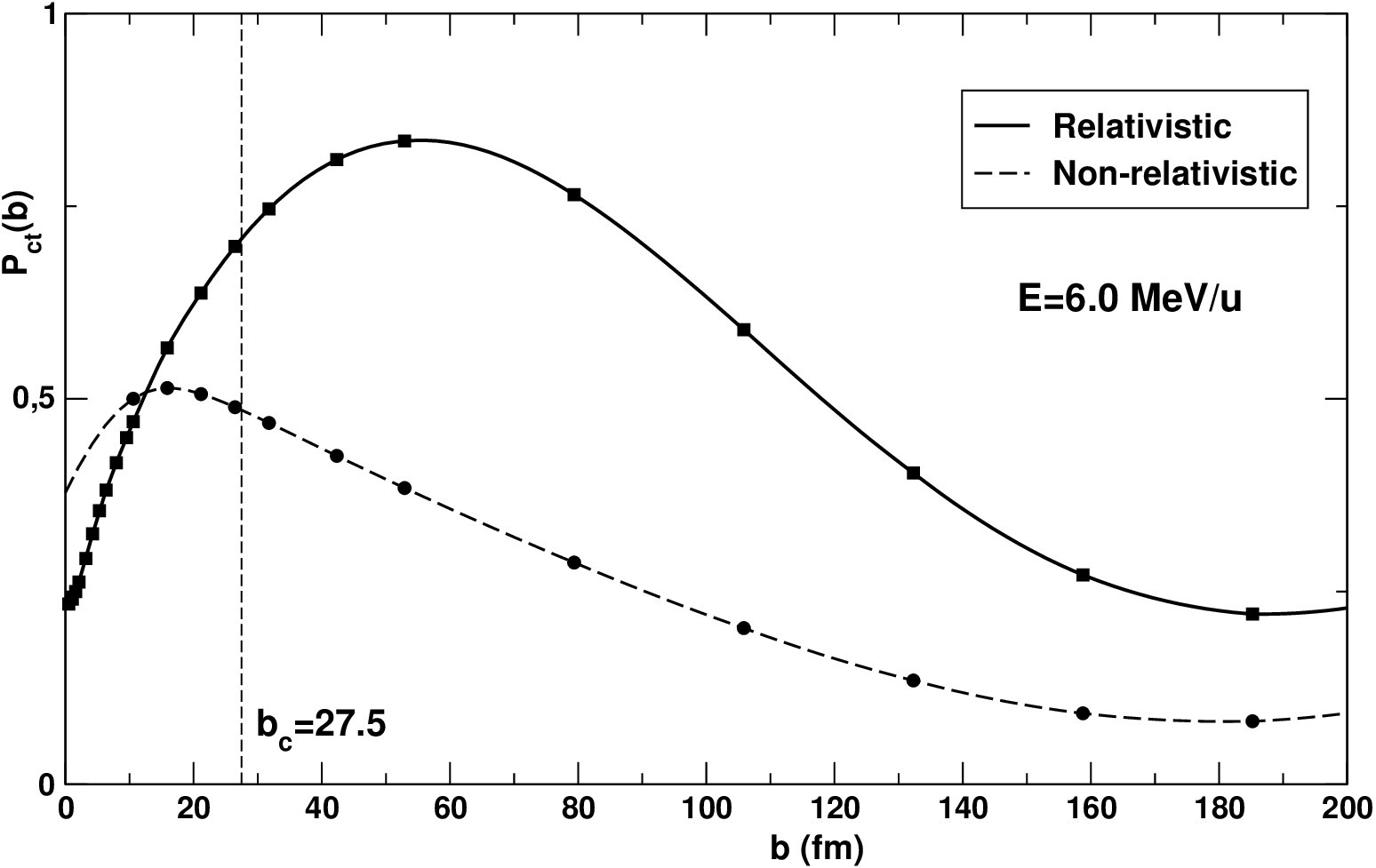}
\caption {Charge transfer probability  P$_{\rm ct}(b)$
 for the U$^{91+}$($1s$) -- U$^{92+}$ collision as a function of 
the impact parameter $b$ in the small $b$ region.
The value $b=b_{\rm c}$  corresponds to the diving  of the  $1\sigma_g$ level
into the negative-energy Dirac continuum. The solid line interpolates the
relativistic values while the dashed line corresponds to the nonrelativistic
limit.}
\label{U2_prob_short}
\end{figure}

The computed relativistic (squares) and non-relativistic (circles) values
of the charge-transfer probabilities $P_{\rm ct}(b)$ and the 
interpolating curves 
are displayed in Fig.~\ref{U2_prob}. It is seen from
the figure, the  nonrelativistic and relativistic probabilities significantly
differ. The nonrelativistic curve (dashed line) is shifted toward higher
energies compared to the relativistic one (solid line).
The same  curves in the
small impact parameter region
are shown in Fig.~\ref{U2_prob_short}.
In this figure, the vertical dashed line indicates the critical impact
parameter $b_{\rm c}=27.5$~fm. For $b=b_{\rm c}$
the  1$\sigma_g$ ground state level of the U$_2^{193+}$
quasi-molecule reaches the negative-energy Dirac continuum.
It should be noted that for the non-straight-line (Rutherford) trajectory
the value of the critical impact parameter $b_{\rm c}$ is less
than the critical distance $R_{\rm c}$ presented in Table~\ref{crit_rad}.
For values $b$ smaller than $b_c$ the 1$\sigma_g$ level  ``dives''
into the negative-energy continuum.

In Table~\ref{U2_cross} we present the results of our relativistic
(3-rd column) and non-relativistic (4-th column) calculations
 of the total
charge-transfer cross section $\sigma_{\rm ct}(E)$, scaled to $Z=1$,
for the  U$^{91+}$($1s$) -- U$^{92+}$ collision at different values of the 
projectile energy $E$.
\begin{table}[ht!]
\small
\caption
{Charge transfer cross section $\sigma_{\rm ct}(E)$
($10^{-17}$ cm$^2$) as a function of the projectile energy $E$
for the U$^{91+}$($1s$)-U$^{92+}$ and H($1s$)-H$^{+}$ collisions.}
\vspace{-3.0mm}
\begin{center}
%
\begin{tabular}{|c|c|c|c|c|c|}
\hline
\multicolumn{1}{|c|}{} & \multicolumn{4}{|c|}{} & 
\multicolumn{1}{c|}{}\\[-6mm]
\multicolumn{1}{|c|}{} &
\multicolumn{4}{|c|}{U$^{91+}$($1s$)-U$^{92+}$} &
\multicolumn{1}{c|}{~H($1s$)-H$^{+}$~}\\[0mm]
\hline &&&&& \\[-6.5mm]
Energy &  Energy & ~~~~$\sigma_{\rm ct} \cdot Z^2$~~~~
       & ~~~~$\sigma_{\rm ct} \cdot Z^2$~~~~
       & ~~~~$\sigma_{\rm ct} \cdot Z^2$~~~~
       & ~~~~$\sigma_{\rm ct}$~~~~  \\[-1mm]
$E/Z^2$ (keV/u)~& ~$E$ (MeV/u)~ & Rel. & Nonrel. &
~Nonrel.~str.~line~ &  \\[1mm]
\hline &&&&&\\[-5.5mm]
 0.70889  &  6.0  & 135.3  &  184.2  &  185.0  &  186.4   \\[-1mm]
 0.76796  &  6.5  & 132.7  &  181.3  &  182.0  &  183.1   \\[-1mm]
 0.82703  &  7.0  & 130.3  &  178.2  &  179.1  &  180.1   \\[-1mm]
 1.18147  & 10.0  & 117.1  &  165.8  &  166.7  &  167.6   \\[1mm]
\hline
\end{tabular}
%
%
\end{center}
\label{U2_cross}
\end{table}
The values of $\sigma_{\rm ct}(E)$ were obtained for
the Rutherford trajectories of the target and projectile ions.
One can see from the table, the relativistic
effect amounts to about $30\%$ of
the non-relativistic value of $\sigma_{\rm ct}$. In the 5-th column of
Table~\ref{U2_cross}, we also present our results obtained for
the straight-line trajectory of the projectile ion
(in this case the target ion is at rest). As one can see from the table,
the difference between the results obtained for the
 straight-line trajectory and
the Rutherford one is very small. The non-relativistic
values of the charge-transfer cross section for the 
 U$^{91+}$($1s$)+U$^{92+}$ collision, scaled to $Z=1$, are also  compared
with the cross section $\sigma_{\rm ct}(E)$ for the H($1s$)-H$^{+}$ collision, 
presented in the 6-th column of Table~\ref{U2_cross}. 
%
\section{Conclusion}
\label{sec:summary}
In this paper we presented a new method for the relativistic
calculations of one-electron two-center quasi-molecular system in
both stationary and time-dependent regimes.
The method is suitable for a wide range of the internuclear distances including
the critical
regime, when the ground state of the quasi-molecule can dive into the
negative-energy Dirac continuum. 
Using this method we calculated the energies of the
H$_2$, Th$_2^{179+}$ and U$_2^{183+}$ quasi-molecules, the critical
distances for some homonuclear quasi-molecules A$^{+(2Z-1)}$
($Z$=$88$, $90$, $92$, $94$, $96$, $98$), and the charge transfer probabilities,
charge
transfer and ionization cross sections for the
H($1s$)--H$^{+}$, Ne$^{9+}$($1s$)--Ne$^{10+}$, Xe$^{53+}$($1s$)--Xe$^{54+}$, and
U$^{91+}$($1s$)--U$^{92+}$ low-energy collisions.

The results of our calculations of
the charge transfer probabilities and cross sections for the
H(1s)--H$^{+}$ collision are in a good agreement with 
experimental data and with theoretical results obtained by other authors.
The influence of the relativistic effect on the charge
transfer probabilities and cross sections for the
Ne$^{9+}$($1s$)--Ne$^{10+}$, Xe$^{53+}$($1s$)--Xe$^{54+}$, and
U$^{91+}$($1s$)--U$^{92+}$ collisions is investigated. We demonstrated, that
the relativistic and nonrelativistic charge-transfer probabilities
as functions of the impact parameter have the same oscillatory behavior
at low energies, but the relativistic curve shifted to lower energies
compared to the nonrelativistic one.
In the case of the U$^{91+}$($1s$)--U$^{92+}$
collision the relativistic effect reduces the values of the cross section by
about $30\%$.

In our further investigations we plan to study in more details the effect of
diving the $1\sigma_g$
level of the U$_{2}^{183+}$ quasi-molecule into the negative-energy Dirac
spectrum and
the influence of this effect on the values of the charge-transfer 
probability. 
With this goal, we are going to develop an
approach
which would allow us to compare the calculated
probabilities
with and without the diving of the ground state into the negative-energy
continuum. We also plan to extend
our method to collisions involving many-electron ions and neutral atoms.
This will allow us to study the $1s$-$1s$ charge transfer in low-energy heavy
ion-atom collisions. Such experiments, that were successfully performed for low-
and middle-$Z$
collisions \cite{Hagmann_82,Hagmann_87,Schuch_84,Schuch_88},
are presently under preparation for high-$Z$ collisions at GSI.
\acknowledgments
We thank V.~Shevelko for providing us with his PWB results for the
charge-transfer probabilities and for valuable discussions.
This work was supported by DFG (Grants No. 436RUS113/950/0-1 and VO 1707/1-1),
by GSI, by RFBR (Grants No. 08-02-91967 and 10-02-00450), by the Ministry of
Education and Science of Russian Federation (Program for Development of
Scientific Potential of High School, Grant No. 2.1.1/1136; Program
``Scientific and pedagogical specialists for innovative Russia'',
Grant No. P1334), and by the ExtreMe Matter Institute EMMI in the
framework of the Helmholtz Alliance HA216/EMMI. 
Y.S.K. acknowledges financial support by the Dynasty Foundation and DAAD. 
V.M.S. acknowledges financial support by the Alexander von Humboldt Foundation.
%
\pagebreak
\clearpage
%

%

\begin{thebibliography}{99}
%
\bibitem{Firsov_51} O.B. Firsov, Zh. Eksp. Teor. Fiz. 
                    {\bf 21}, 1001 (1951).
%
\bibitem{Demkov_52} Yu.N. Demkov, Uchen. Zap. Leningr. Univ. 
                    {\bf 146}, 74 (1952).
%
\bibitem{Bates_53} D.R. Bates, H.S.W. Massey, and A.L. Stewart,
                    Proc. Roy. Soc (London) A{\bf 216}, 437 (1953).
%
\bibitem{Bransden_92}  B.H. Bransden and M.R.C. Mcdowell,
                  \textit{Charge Exchange and the Theory of Ion-Atom
Collisions}, (Oxford University Press, NY 1992).
%
\bibitem{Fritsch_91} W. Fritsch and C.D. Lin, Phys. Rep. {\bf 202}, 1 (1991).
%
\bibitem{Winter_05} T.G. Winter, Adv. At. Mol. Opt. Phys. {\bf 52},
                    391 (2005).
%
\bibitem{Gallaher_68} D.F. Gallaher, L. Wilets, Phys. Rev.
                     {\bf 169}, 139 (1968).
%
\bibitem {Shakeshaft_76} R. Shakeshaft, Phys. Rev. A {\bf 14}, 1626 (1976).
%
\bibitem {Reading_81} J.F. Reading, A.L. Ford and R.L. Becker,
                      J. Phys. B {\bf 14} 1995 (1981).
%
\bibitem {Fritsch_83} W. Fritsch, C.D. Lin, Phys. Rev. A {\bf 27},
                      3361 (1983).
%
\bibitem {Ermolaev_90} A.M. Ermolaev, J. Phys. B {\bf 23}, L45 (1990).
%
\bibitem {Toshima_99} N. Toshima, Phys. Rev. {\bf 59}, 1981 (1999).
%
\bibitem{Winter_09} T.G. Winter, Phys. Rev. A {\bf 80}, 032701 (2009).
%
\bibitem{Grun_82} N. Gr\"un, A. M\"uhlhans and W. Scheid,
                  J. Phys. B {\bf 15}, 4043 (1982).
%
\bibitem{Kolakowska_98} A. Kolakowska, M.S. Pindzola, F. Robicheaux,
                  D.R. Schultz, C. Wells, Phys. Rev. A {\bf 98}, 2872 (1998).
%
\bibitem{Kolakowska_99} A. Kolakowska and M.S. Pindzola, D.R. Schultz,
                        Phys. Rev. A {\bf 59}, 3588 (1999). 
%
\bibitem{Schultz_99} D.R. Schultz, M.R. Strayer and J.C. Wells,
                     Phys. Rev. Lett. {\bf 82}, 3976 (1999).
%
\bibitem{Tong_00} X. Tong, D. Kato, T. Watanabe, and S. Ohtani
                  Phys. Rev. A {\bf 62}, 052701 (2000).
%
\bibitem{Briggs_73} J.S. Briggs, and J.H. Macek, 
                    J. Phys. B {\bf 6}, 982 (1973);
                    Corrigenda {\bf 6}, 2484 (1973).
%
\bibitem{Eichler_95} J. Eichler and W.E. Meyerhof, \textit{Relativistic Atomic
                     Collisions}, (Academic, New York, 1995).
%
\bibitem{Shabaev_02} V.M. Shabaev, Phys. Rep. {\bf 356}, 119 (2002).

%
\bibitem{Eichler_07} J. Eichler and T. St\"ohlker, Phys. Rep. {\bf 439}, 1
                     (2007).
%
\bibitem{Zeldovich_71} Y.B. Zeldovich and V.S. Popov, Usp. Fiz. Nauk
                       {\bf 105}, 403 (1971)
                       [Sov. Phys. Usp. {\bf 14}, 673 (1972)].
%
\bibitem{Rafelski_71} J. Rafelski, L.P. Fulcher, and W. Greiner,
                      Phys. Rev. Lett. {\bf 27}, 958 (1971).
%
\bibitem{Greiner_85}  W. Greiner, B. M\"uller, J. Rafelski,
                 \textit{Quantum Electrodynamics of Strong Fields},
                 (Springer-Verlag, Berlin, 1985).
%
\bibitem{Muller_94} U. M\"uller-Nehler and G. Soff, Phys. Rep.
                    {\bf 246}, 101 (1994).
%
\bibitem{Rafelski_76}  J. Rafelski, B. M\"uller, Phys. Lett.
                       {\bf 65}B, 205 (1976).

%
\bibitem{Eichler_05} J. Eichler, \textit{Lectures on Ion-Atom Collisions: From Nonrelativistic
to Relativistic Collisions}, (Elsevier, Amsterdam, 2005).

%
\bibitem{Becker_86} U. Becker, N. Gr\"un, W. Scheid, and G. Soff,
                      Phys. Rev. Lett. {\bf 56}, 2016 (1986).
%
\bibitem{Strayer_90} M.R. Strayer, C. Bottcher, V.E. Oberacker, and A.S. Umar,
                     Phys. Rev. A {\bf 41}, 1399 (1990).
%
\bibitem{Thiel_92} J. Thiel, A. Bunker, K. Momberger, N. Gr\"un, and W. Scheid,
                   Phys. Rev. A {\bf 46}, 2607, (1992).
%
\bibitem{Momberger_96} K. Momberger, A. Belkacem, and A.H. Sorensen,
                       Phys. Rev. A {\bf 53}, 1605  (1996).
%
\bibitem{Wells_96} J.C. Wells, V.E. Oberacker, M.R. Strayer, and A.S. Umar,
                   Phys. Rev. A {\bf 53}, 1498 (1996).
%
\bibitem{Ionescu_99} D.C. Ionescu and A. Belkacem, Physica Scripta {\bf T80},
                     128 (1999).
%
\bibitem{Pindzola_00} M.S. Pindzola, Phys. Rev. A {\bf 62}, 032707 (2000).
%
\bibitem{Busic_04} O. Busic, N. Gr\"un, and W. Scheid,
                   Phys. Rev. A {\bf 70}, 062707 (2004).

\bibitem{Eichler_90} J. Eichler, Phys. Rep. {\bf 193},  165 (1990).

%
\bibitem{Rumrich_93} K. Rumrich, G. Soff, W. Greiner, Phys. Rev. A {\bf 47},
                     215 (1993).
%
\bibitem{Momberger_93} K. Momberger, N. Gr\"un, and W. Scheid,
                       J. Phys. B {\bf 26}, 1851 (1993).
%
\bibitem{Gail_03} M. Gail, N. Gr\"un and W. Scheid, J. Phys. B
                  {\bf 36}, 1397 (2003).
%
\bibitem{Soff_78} G. Soff, J. Reinhardt, and W. Betz, Phys. Scr.
                  {\bf 17}, 417 (1978).
%
\bibitem{Reus_84} T.H.J de Reus, J. Reinhardtt, B. M\"uller, W. Greiner,
         G. Soff, and U. M\"uller, J. Phys. B {\bf 17}, 615 (1984).
%
\bibitem{Ackad_08} E. Ackad and M. Horbatsch, Phys. Rev. A {\bf 78},
                   062711 (2008).
%
\bibitem{Stohlker_98a} Th. St\"ohlker, D.C. Ionescu, P. Rymuza, F. Bosch,
                      H. Geissel, C. Kozhuharov, T. Ludziejewski,
                      P.H. Mokler, C. Scheidenberger, Z. Stachura,
                      A. Warczak and R. W. Dunford,
                      Phys. Lett. A {\bf 238}, 43 (1998).
%
\bibitem{Stohlker_98b} Th. St\"ohlker, D.C. Ionescu, P. Rymuza, Z. Stachura,
                       A. Warczak, and R. W. Dunford,
                       Phys. Rev. A {\bf 57}, 845 (1998).
%
\bibitem{Ionescu_03} D.C. Ionescu and Th. St\"ohlker, Phys. Rev.
                    {\bf 67}, 022705 (2003).

%
\bibitem{Rotenberg_70} M. Rotenberg, Adv. At. and Mol. Phys. {\bf 6},
                       233 (1970).
%
\bibitem{Gruzdev_87} P.F. Gruzdev, G.S. Soloveva, A.I. Sherstyuk,
                     Opt. and Spectrosc, {\bf 63}, 1394 (1987),
%
\bibitem{Manakov_73} N.L. Manakov, L.P. Rapoport and S.A. Zapryagaev,
                     Phys. Lett. A {\bf 43}, 139 (1973).
%
\bibitem{Drake_88}   G.W.F. Drake and S.P. Goldman, Adv. At. Mol. Phys.
                     {\bf 25}, 393 (1988).
%
\bibitem{Avery_98}   J. Avery and F. Antonsen,  J. Math. Chem. {\bf 24},
                      175 (1998).
%
\bibitem{Grant_00}   I.P. Grant and H.M. Quiney, Phys. Rev. A {\bf 62},
                     022508 (2000).
%
\bibitem{Tupitsyn_03} I.I. Tupitsyn, V.M. Shabaev, J.R. Crespo L\'opez-Urrutia,
                      I. Draganic, R. Soria Orts, and J. Ulrich,
                      Phys. Rev. A {\bf 68}, 022511 (2003).
%
\bibitem{Tupitsyn_05} I.I. Tupitsyn, A.V. Volotka, D.A. Glazov, V.M. Shabaev,
                      G.Plunien, J.R.Crespo Lopez-Urrutia, A.Lapierre,
                      and J. Ullrich,  Phys. Rev. A {\bf 72}, 062503 (2005).
%
\bibitem{Lowdin_56}  P.O. L\"owdin, Adv. in Phys. {\bf 5}, 1 (1956).
%
\bibitem{Kotochigova_95} S. Kotochigova, I. Tupitsyn,
                         Int. J. Quantum Chem. {\bf 29}, 307 (1995).
%
\bibitem{Tupitsyn_98} I.I. Tupitsyn, D.A. Savin, and V.G. Kuznetsov,
                   Opt. and Spectrosc. {\bf 84}, 344 (1998).
%
\bibitem{Yang_91}  L. Yang, D. Heinemann, D. Kolb, Chem. Phys. Lett.
                   {\bf 178}, 213 (1991).
%
\bibitem{Kullie_01} O. Kullie, D. Kolb, Eur. Phys. J. D {\bf 17},
                    167 (2001).
%
\bibitem{Lisin_77} V.I. Lisin, M.S. Marinov, V.S. Popov, Phys. Lett.
                    {\bf 69}B, 2 (1977).
%
\bibitem{Matveev_00} V.I. Matveev, D.U. Matrasulov, and H.Yu. Rakhimov,
                     Phys. At. Nucl. {\bf 63}, 318. (2000).
%
\bibitem{Muller_76} B. M\"uller, and W. Greiner, Z. Naturforsch.
                    {\bf 31}a, 1 (1976).
%
\bibitem{Lisin_80} V.I. Lisin, M.S. Marinov, V.S. Popov, Phys.Lett.
                    {\bf 91}B, 20 (1980).
%
\bibitem{Popov_01} V.S. Popov, Phys. At. Nucl. {\bf 64},
                   367, (2001).
%
\bibitem{Crank_47} J. Crank, P. Nicholson, Proc. Cambridge Philos. Soc.
                 {\bf 43}, 50 (1947).
%
\bibitem{Fett_82} M.D. Feit, J.A. Fleck, Jr., and A. Steiger,
                 J. Comput. Phys. {\bf 47}, 412 (1982).
%
\bibitem{Wilkinson_71} J. Wilkinson, C. Reinsch, \textit{Handbook
     for Automatic Computation}, (Springer-Verlag, Berlin, 1971),
%
\bibitem{Grant_91} I.P. Grant, B.L. Gyorffy, \textit{The Effects of Relativity
                in Atoms, Molecules, and the Solid State}, (ed. S Wilson,
                New York: Plenum, 1991).
%
\bibitem{Szmytkowski_97} R. Szmytkowski, J. Phys. B {\bf 30}, 825 (1997).
%
\bibitem{Grant_70}  I.P. Grant, Adv. Phys. {\bf 19}, 747 (1970).
%
\bibitem{Bratsev_77} V.F. Bratsev, G.B. Deyneka, and I.I. Tupitsyn,
                     Bull.Acad.Sci. USSR, Phys. Ser. {\bf 41},
                     173 (1977).
%
\bibitem{Sharma_76} R.R. Sharma, Phys. Rev. A {\bf 13}, 517 (1977).
%
\bibitem{Rose_61} M.E. Rose,\textit{ Relativistic Electron Theory},
                  (John-Wiley \& Sons, NY-London, 1961)
%
\bibitem{Varshalovich_88} D.A. Varshalovich, A.N. Moskalev,
              V.K. Khersonskii, \textit{Quantum Theory of Angular Momentum},
              (World Scientific, Singapore, 1988).
%
\bibitem{Marsden_03} J.E. Marsden, A.J. Tromba, \textit{Vector Calculus},
                     (5th edition, W.H. Freeman \& Company, New-York,
                    2003).
%
\bibitem{Condon_35} E.U. Condon, G.H. Shortley, \textit{Theory of Atomic
Spectra},
                     (Cambridge University Press, London, 1935).
%
\bibitem{Shab_Tup_04} V.M. Shabaev, I.I. Tupitsyn, V.A. Yerokhin,
                      G. Plunien, and G. Soff, Phys. Rev. Lett.
                      {\bf 93}, 130405 (2004).
%
\bibitem{Tup_Shab_08} I.I. Tupitsyn, V.M. Shabaev,
               Opt. and Spectrosc. {\bf 105}, 183 (2004).
%
\bibitem{Parpia_95} F.A. Parpia, A.K. Mohanty, Chem. Phys. Lett.
                    {\bf 238}, 209 (1995).

\bibitem{LaJohn_98} L. LaJohn, J.D. Talman, Theor. Chim. Acta.
                    {\bf 99}, 351 (1998).
%
\bibitem{Rutkowski_99} A. Rutkowski, Chem. Phys. Lett. {\bf 307}, 259 (1999).
%
\bibitem{Muller_72} B. M\"uller, J. Rafelski, and W. Greiner, Z. Physik
                     {\bf 257}, 183 (1972).
%
\bibitem{Marinov_75} M.S. Marinov, V.S. Popov, and V.L. Stolin,
                     J. Comp. Phys. {\bf 19}, 241 (1975).
%
\bibitem{Wietschorke_79} K-H Wietschorke, B. Muller, W. Greiner and G. Soff,
                      J. Phys. B {\bf 12}, L31 (1979).
%
\bibitem{Parpia_92} F.A. Parpia, A.K. Mohanty, Phys. Rev. A {\bf 46},
                    3735 (1992).
%
\bibitem{Angeli_04} I. Angeli, At. Data Nucl. Data Tables {\bf 87},
                    185 (2004).
%
\bibitem{Kozhedub-Shabaev_08}  Y.S. Kozhedub, V.M. Shabaev,
                               unpublished.
%
\bibitem{Kozhedub_08}  Y.S. Kozhedub, O.V. Andreev, V.M. Shabaev,
       I.I. Tupitsyn, C. Brandau, C. Kozhuharov, G. Plunien, and
       T. St\"ohlker, Phys. Rev. A {\bf 77}, 032501 (2008).
%
\bibitem{Johnson-Soff_85} W.R. Johnson, G. Soff, At. Data Nucl. Data Tables
                    {\bf 33}, 405 (1985).
%
\bibitem{Fritsch_82} W. Fritsch and C.D. Lin, Phys. Rev. A
                    {\bf 26}, 762 (1982).
%
\bibitem{Ludde_81} H.J. L\"udde and R.M. Dreizler, J. Phys. B 
                   {\bf 14}, 2191 (1981).
%
\bibitem{Ludde_82} H.J. L\"udde and R.M. Dreizler. J. Phys. B
                   {\bf 15}, 2703 (1982).
%
\bibitem {Shakeshaft_78} R. Shakeshaft, Phys. Rev. A {\bf 18}, 1930 (1978).
%
\bibitem{Janev_88} R.K. Janev, J.J. Smith, Atomic and Plasma Material
       Interaction Data for Fusion, Nucl Fusion Suppl. Special Issue,
       {\bf 4} (1993).
%
\bibitem{Shah_87} M.B. Shah J. Phys. B {\bf 20}, 2481 (1987).
%
\bibitem{Shah_98} M.B. Shah J. Phys. B {\bf 31}, L757 (1998).
%
\bibitem{Shevelko_09} V.P. Shevelko, private communication.
%
\bibitem{Shevelko_01} V.P. Shevelko, I.Yu. Tolstikhina, Th. St\"ohlker,
                      Nucl. Instrum. Methods B {\bf 184}, 295 (2001).

%
\bibitem{Hagmann_82} S. Hagmann, C. L. Cocke, J. R. Macdonald, P.
Richard,  H. Schmidt-B\"ocking, and R. Schuch, Phys. Rev. A
                    {\bf 25}, 1918 (1982).
%
\bibitem{Hagmann_87} S. Hagmann, S. Kelbch, H. Schmidt-B\"ocking,  C. L. Cocke,
P. Richard, R. Schuch, A. Skutlartz,   J. Ullrich,   B. Johnson, M. Meron, K.
Jones,  D. Trautmann and F. R\"osel,  Phys. Rev. A
                    {\bf 36}, 2603 (1987).
%
\bibitem{Schuch_84} R. Schuch, H. Ingwersen, E. Justiniano, H. SchmidtBocking, M.
Schulz, and F. Ziegler, J. Phys. B {\bf 17}, 2319 (1984).

%
\bibitem{Schuch_88} R. Schuch, M. Meron, B. M. Johnson, K. W. Jones, R.
Hoffmann,  H. Schmidt-B\"ocking,  I. Tserruya,     Phys. Rev. A
                    {\bf 37}, 3313 (1988).
%

\end{thebibliography}
\end {document}